\documentclass[%
 reprint,
 amsmath,amssymb,
 aps,
]{revtex4-2}

\usepackage{graphicx}
\usepackage{dcolumn}
\usepackage{bm}
\usepackage{amsmath}

\begin{document}

\preprint{APS/123-QED}

\title{Bottom-up data integration in polymer models of chromatin organisation}

\author{Alex Chen Yi Zhang}
 \email{chzhang@sissa.it}
\author{Angelo Rosa}
 \email{anrosa@sissa.it}
\author{Guido Sanguinetti}
 \email{gsanguin@sissa.it}
\affiliation{%
  Scuola Internazionale Superiore di Studi Avanzati (SISSA), Via Bonomea 265, 34136 Trieste, Italy
}%

\date{\today}


\begin{abstract}
Cellular functions crucially depend on the precise execution of complex biochemical reactions taking place on the chromatin fiber in the tightly packed environment of the cell nucleus.
Despite the availability of large data sets probing this process from multiple angles, we still lack a bottom-up framework which can incorporate the sequence-specific nature of biochemistry in a unified model of 3D chromatin dynamics.
Here we propose {\it SEMPER} ({\it Sequence Enhanced Magnetic PolymER}), a novel stochastic polymer model which naturally incorporates observational data about sequence-driven biochemical processes, such as binding of transcription factor proteins, in a 3D model of chromatin structure.
By introducing a new algorithm for approximate Bayesian inference, we discuss how to estimate in a robust manner the relative importance of biochemical {\it vs.} polymer signals in the determination of the chromatin epigenetic states which is leading to a significant revision of the interpretation of previous models. 
Furthermore we show that, without additional input from the genome 3D structure, our model can predict with reasonable accuracy some notable and non trivial conformational features of chromatin folding within the nucleus. 
Our work highlights the importance of introducing physically realistic statistical models for predicting chromatin states from epigenetic data, and opens the way to a new class of more systematic approaches to interpret epigenomic data.
\end{abstract}

\maketitle


\section{Introduction}\label{sec:Intro}
Genomes' tight confinement~\cite{SazerSchiessel-2018} within the nucleus of the cell during interphase
plays a crucial role in many fundamental biological processes {\it in vivo}, ranging from gene expression and gene regulation to DNA replication~\cite{RowleyCorces2018,FinnMisteli2019,Misteli2020}.
However, how such mechanical constraints on the chromatin fibers interact with the genetic signals which regulate cellular biology at the molecular level, and how the genome $1D$ sequence is integrated within the complex $3D$ nuclear environment to determine the {\it epigenetic} fate of each cell, 
remain amongst the most challenging open problems of modern biology~\cite{Merlotti2020,Misteli2020}. 

The chromatin filament of each chromosome is organized into distinct motifs that characterize genome folding at different scales in $3D$:
topologically-associated domains (TAD's~\cite{Dixon2012}),
nuclear (A/B) sub-compartments~\cite{Dekker2009}
and
chromosome territories~\cite{CremerBros2001}
all participate to genome regulation.
Quite surprisingly, the molecular origins of these structures appear remarkably ``simple'', having been explained as the consequence of general physical mechanisms based on polymer physics, such as
molecular ``bridges''~\cite{Nicodemi2012,Brackley2016},
polymer microphase separation~\cite{Jost2014,Safran2021},
DNA-loop extrusion~\cite{Marko2012,FudenbergMirny2016}
and
topological constraints~\cite{RosaPlos2008,Halverson2014}.
On the other hand, cells need to execute genetic programs in a robust and coordinated fashion which requires the efficient transcriptional activation of specific, and cell-type dependent, genomic regions, a fact that cannot be accounted for only by general physical principles.
So far, the biochemical bases of these processes have been explained solely, yet with remarkable success, in terms of the $1D$ structure of the genome, particularly in terms of binding of sequence-specific transcription factor (TF) proteins~\cite{ptashne2014chemistry,gann2002genes} and other epigenetic marks, such as covalent modifications of histone tails~\cite{AlbertsMBC}, to the underlying DNA sequence.
In this respect, 
there is now a mature line of research showing how machine learning (ML) algorithms trained on sequence-related features can be used to explain and predict the transcriptional as well as the epigenetic states of the cells~\cite{Benveniste2014, whitaker2015predicting, schreiber2020avocado,hawkins2022getting}.
Interestingly, the success of these ML tools suggests that their systematic integration into polymer-based approaches should, potentially, pave the way for a viable strategy to reach a more complete understanding of genome folding in $3D$ based on the underlying $1D$ sequence.

In order to explore such possibility, we introduce here {\it SEMPER (Sequence Enhanced Magnetic PolymER)}, a new statistical model for chromosome conformations which explicitly bridges between the $1D$ sequence-informed structure of the genome and its $3D$ nature as a polymer. 
It does so by {\it interpolating} between a spatial magnetic polymer model~\cite{ColiMichielettoPRE2019,MichielettoColiPRL2019} describing domains formation in chromatin fibers and a 
statistical predictor~\cite{Benveniste2014} of epigenetic states based on TF-binding data. 
As, however, a direct Bayesian estimation of the parameters introduced in the model is intractable, we develop here a novel approximate inference method to estimate the model parameters and quantify the corresponding uncertainties.
We apply this model to study the chromosomal organisation of selected sequences from human chromosome 2 at different levels of resolution.
We compare our results in terms of chromatin state predictions with a purely data-driven benchmark~\cite{Benveniste2014}, leading to both improved predictions and a substantial reappraisal of the role of the various molecular factors in determining the chromatin state.

Additionally, and as a by-product of our methodology, {\it SEMPER} explicitly provides putative chromatin conformations, hence it can be used to explore the specific patterns that emerge in the chromatin contact maps obtained by Hi-C technologies~\cite{Dekker2009,Dixon2012}.
Overall, our work demonstrates that a suitable bottom-up polymer model enhanced with a relatively {\it small} number of sequence ``features'' can both reproduce the epigenetic state of the chromosome and its local $3D$ organisation, providing a new tool to conceptualise biochemistry and gene regulation at the microscopic level.

The paper is organized as follows.
In Section~\ref{sec:Methods} we summarize the details of our computational model, in particular the model Hamiltonian and the physical meaning of the distinct terms and parameters which define it (Sec.~\ref{sec:ModelHamiltonian}) as well as the (approximate) Monte Carlo procedure used to estimate the parameters and the relative uncertainties (Sec.~\ref{sec:EstimatingParams}).
In Sec.~\ref{sec:Results}, we demonstrate the validity and generality of our approach by focusing on some specific sequences of the human chromosome 2:
after a description of the data sets used for the model (Sec.~\ref{sec:FeatureSelection}), we discuss the predictions of our model for the epigenetic state of the chromatin fiber (Sec.~\ref{sec:EpiStatesPrediction}) and the role that the polymeric nature of chromatin plays in it. Then (Sec.~\ref{sec:BottomUpAndInteractions}), we measure the chromatin contacts within our model conformations and compare them to the corresponding experimental Hi-C matrices.
Finally (Sec.~\ref{sec:DiscConcls}), we conclude by discussing the main advantages of our model with respect to other approaches and point out future possible improvements.
In Appendices~\ref{app:simulations} and~\ref{app:approximate_mcmc} the interested reader will find more details about the polymer model and the simulation scheme, while in Appendix~\ref{app:figs} we present additional figures illustrating the discretization of the ChIP-seq tracks, the convergence of the method for synthetic data and a summary of the {\it ENCODE} accession codes relative to the biological data used as inputs of our model.

\begin{figure*}
\centering
\includegraphics[width=0.8\linewidth]{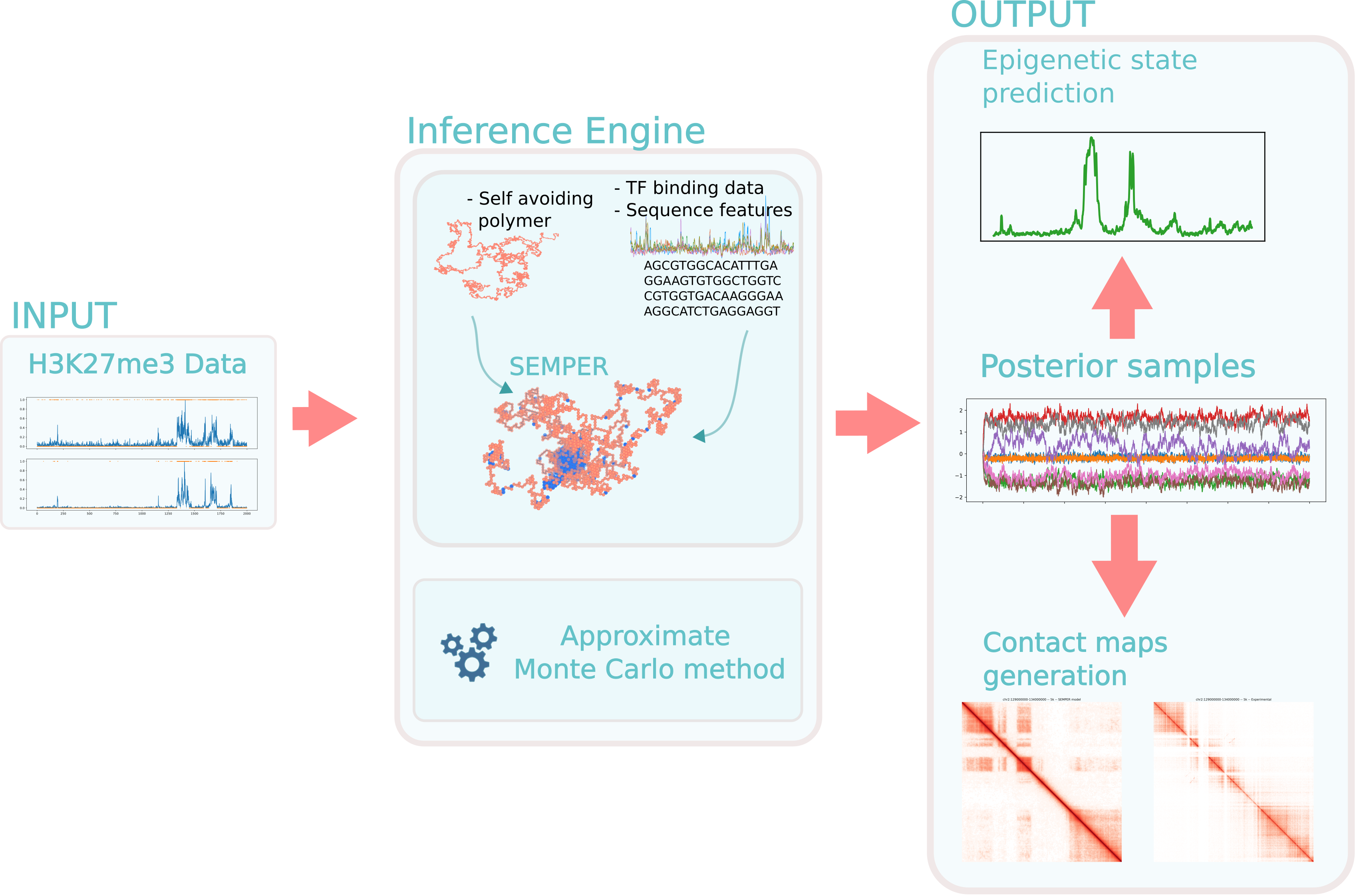}
\caption{\label{fig:magnetic_polymer}
Illustration of the {\it SEMPER} numerical pipeline.
The model integrates the epigenetic content of the $1D$ DNA sequence within a $3D$ polymer model for the chromatin fiber. 
Starting from {\it H3K27me3} methylation data our approximate Metropolis-Hastings Monte Carlo method 
allows to perform posterior inference on the otherwise intractable model.
The obtained posterior samples can then be used to predict both, the \emph{1D} epigenetic state and the \emph{3D} folding of the chromatin fiber extracted from the Hi-C contact maps.
Notably the model gets no input from $3D$ chromatin folding, so it can be used to get insight on chromatin conformations based only on the information acquired from the sequence. 
}
\end{figure*}
%

\section{Polymer model and Monte Carlo methods, materials}\label{sec:Methods}

\subsection{An integrative model of chromatin organisation}\label{sec:ModelHamiltonian}
Eukaryotic genomes are organized in single chromosomes, and each chromosome is made of a unique filament of chromatin fiber.
Chromatin itself is a sophisticated filamentous DNA-protein complex which serves as a platform for the integration of multiple biochemical and biophysical signals~\cite{AlbertsMBC}. 

A convenient model to capture the $3D$ organisation of the chromatin is the {\it magnetic polymer model}~\cite{GarelOrlandOrlandini1999, ColiMichielettoPRE2019, MichielettoColiPRL2019}: a polymer chain in $3D$ is modelled as a self-avoiding random walk (SAW), with nearby (in space) monomers interacting via a Ising-like term.
While this model has been shown to recapitulate successfully the statistics of open/closed chromatin domains, it does not account for the dynamical activation of specific chromatin loci as a result of sequence-specific biochemical signals.
To address this issue, we expand the model by introducing monomer-specific {\it external field} terms, which modulate the probability of each monomer lying in an open or closed region based on its biochemical features, such as the sequence composition or the presence of specific binding sites for transcription factors (Fig.~\ref{fig:magnetic_polymer}).

More precisely (see Appendix~\ref{app:simulations} for details), chromosome spatial conformations are modeled as $N$-steps self-avoiding paths on the cubic lattice~\cite{Madras:1988ei} of linear step $=d$.
For computational convenience we work with polymer chains with $N=1000$, while we vary $d$ systematically in order to study the chromatin fiber at different resolution details (see Table~\ref{tab:pearson_corr_histone}).

Then, model chromosomal conformations are sampled from the canonical ensemble according to the following Hamiltonian:
\begin{eqnarray}
\label{eq:hamiltonian}
{\mathcal H}(\{r\}, \{\sigma\} | \pmb{\theta})
& = & {\mathcal H}_{\rm SAW}(\{r\})\, -\,\frac{J}{2}\sum_{i\neq j}\sigma_{i}\sigma_{j}\Delta(\pmb{r}_i,\pmb{r}_j) \nonumber\\
& & -\sum_{i}h_i(\pmb{w})\sigma_i \nonumber\\
& \equiv & {\mathcal H}_{\rm SAW}(\{r\}) + {\mathcal H}_{\rm Ising}(\{r\}, \{\sigma\} | J) \nonumber\\
 & & + {\mathcal H}_{\rm ext}(\{\sigma\} | \pmb{w}) \, ,
\end{eqnarray}
where
$\{r\} \equiv ({\mathbf r}_1, ..., {\mathbf r}_N)$
and
$\{\sigma\} \equiv (\sigma_1, ..., \sigma_N)$
are the compact notations for, respectively, the spatial coordinates of the $N$ monomers composing the polymer chain
and
the corresponding Ising-like spin variables which take the two possible values $\sigma_i = 1 / 0$ (accounting for, respectively, the presence/absence of the histone modification, see below).

The three terms in Eq.~\eqref{eq:hamiltonian} represent the following.
$\mathcal{H}_{\rm SAW}(\{r\})$ enforces strict self-avoidance of the polymer chain on the cubic lattice~\cite{Madras:1988ei}: it is equal to $+\infty$ whenever any two monomers occupy the same lattice site and $=0$ otherwise. 
The second term ${\mathcal H}_{\rm Ising}(\{r\}, \{\sigma\} | J)$, with $J>0$ and $\Delta(\pmb{r}_i, \pmb{r}_j) = 1$ for $|\pmb{r}_i - \pmb{r}_j| = d$ and $=0$ otherwise, is the Ising-like interaction generating the magnetic effect.
Finally, the external field term ${\mathcal H}_{\rm ext}(\{\sigma\} | \pmb{w})$ models the coupling of the chromatin state with the underlying biochemistry: we compute the value of the {\it external-field} at each monomer location $i$ as a linear combination of sequence and TF-binding derived features, {\it i.e.} $h_i(\pmb{w}) = \sum_{\alpha=1}^f F_{i\alpha}w_{\alpha}$.
For compactness, we denote by $\pmb{\theta} \equiv (J, \pmb{w})$ the free parameters of the polymer model, namely the spin-spin coupling constant $J$ and the $f$ weights ($\pmb{w} \equiv (w_1, w_2, ..., w_f)$) used in the computation of the external fields.

Notice that in the limit when the magnetic polymer structure is ignored ({\it i.e.}, for $J=0$), the spin variables decouple and
the probability of a particular spin sequence is given by the logistic regression predictor considered in~\cite{Benveniste2014}.
In this way, the polymer model described by~\eqref{eq:hamiltonian} interpolates between a biophysical model based on magnetic polymers and a data-driven predictor for the epigenetic state of the sequence.

\subsection{Estimating model parameters: approximate Monte Carlo method}\label{sec:EstimatingParams} 
The introduction of external fields in~\eqref{eq:hamiltonian} provides additional flexibility to capture local patterns in chromatin organisation, at the cost of introducing a number of additional parameters.
Estimating these parameters can in principle be done in a Bayesian way by employing experimental data for the chromatin state, for instance through ChIP-seq experiments~\cite{ChipSeq2007} detecting histone modifications as a proxy for open/closed chromatin. 

Explicitly, we assume the availability of experimental observations $\{\tilde{\sigma}\}$ for the state of each monomer (open/closed).
Then, by measuring energy in units of $\kappa_B T=1$ where $\kappa_B$ is the Boltzmann constant, the likelihood associated with such a spin configuration at fixed $\pmb{\theta} = (J, \pmb{w})$ reads:
\begin{eqnarray}\label{eq:likelihood}
P(\{\tilde{\sigma}\} | \pmb{\theta})
& = & \sum_{\{r\} \in {\rm SAW}} P(\{r\}, \{\tilde{\sigma}\} |\pmb{\theta}) \nonumber\\
& = & \frac{e^{-{\mathcal H}_{\rm ext}(\{\tilde{\sigma}\} | \pmb{w})}}{Z(\pmb{\theta})} \Phi(J , \{\tilde{\sigma}\}) \, ,
\end{eqnarray}
with
\begin{eqnarray}
Z(\pmb{\theta})
& = & \sum_{\{r\} \in {\rm SAW}} \sum_{\{\sigma\}} \exp\left( -\mathcal{H}(\{r\}, \{\sigma\} | \pmb{\theta}) \right) \, , \label{eq:Ztheta} \\ 
\Phi(J , \{\tilde{\sigma}\})
& = & \sum_{\{r\} \in {\rm SAW}} e^{-\left( {\mathcal H}_{\rm SAW}(\{r\}) + {\mathcal H}_{\rm Ising}(\{r\}, \{\tilde{\sigma}\} | J) \right)} \, , \label{eq:PhiJ}
\end{eqnarray}
where $P(\{r\}, \{\tilde{\sigma}\} | \pmb{\theta})$ is the canonical distribution with the Hamiltonian~\eqref{eq:hamiltonian}.

Unfortunately, the calculation of the likelihood~\eqref{eq:likelihood} entails the marginalisation of the polymer configurations which, except for the $J=0$ case, constitutes an analytically intractable task. Therefore the posterior distribution
%
\begin{equation}\label{eq:PosteriorPThetaSigmaTilde}
    P(\pmb{\theta} | \{\tilde{\sigma}\}) \propto
    P(\{\tilde{\sigma}\} | \pmb{\theta}) P(\pmb{\theta}) =
    \frac{e^{-{\mathcal H}_{\rm ext}( \{\tilde{\sigma}\} | \pmb{w})}}{Z(\pmb{\theta})} \Phi(J, \{\tilde{\sigma}\}) P(\pmb{\theta}) \, ,
\end{equation}
also depends on the intractable normalization constants
$Z(\pmb{\theta})$ and $\Phi(J, \{\tilde{\sigma}\})$, hence exact sampling with standard tools such as the popular Metropolis-Hastings (MH) algorithm \cite{Metropolis1953} is unfeasible. Indeed in the MH Markov Chain Monte Carlo (MCMC) algorithm, given the current state $\pmb{\theta}^t$ of the chain the new state $\pmb{\theta}^{t+1}$ (proposed by adding a random displacement $\pmb{\delta \theta}^{t}$) can, in principle, be accepted with a probability given by the following Metropolis-Hastings ratio:
\begin{equation}\label{eq:acceptance}
   \alpha = \left(1,\,\frac{P(\pmb{\theta}^{t+1}|\{\tilde{\sigma}\})}{P(\pmb{\theta}^t|\{\tilde{\sigma}\})}\right) \, .
\end{equation}
which contains ratios of the said intractable normalization constants.

Various methodologies, collectively known as {\it likelihood-free} or {\it simulator-based} inference methods, have been developed over the years to perform approximate Bayesian inference in settings where the likelihood is not analytically tractable. 
Such methods, recently reviewed in~\cite{beaumont02,cranmer2020frontier}, are principally geared towards situations where the stochastic process can only be simulated.
The error made in these approximations is typically difficult to control and generally, these methods incur very high computational costs particularly when the number of parameters to estimate is from medium to high.
The main cause of both these drawbacks is that typically these methods treat the simulator as black box sample generators, without taking advantage of 
the characteristics of the system at hand.
Here, instead, we 
propose to leverage the description of the system in terms of the Hamiltonian~\eqref{eq:hamiltonian} to enable a more guided exploration of the parameters space, thus providing a more efficient and scalable algorithm.

For this purpose, we propose an approximate Metropolis-Hastings inference method based on the observation that we may take advantage of the following Taylor expansions:
\begin{equation}\label{eq:logZTaylorExp}
\ln{\frac{Z(\pmb{\theta}+\pmb{\delta \theta})}{Z(\pmb{\theta})}} \simeq \pmb{\delta \theta}\cdot\nabla_{\pmb{\theta}} \ln{Z(\pmb{\theta})} + \frac{\pmb{\delta \theta}^T \cdot \pmb{H}_{\ln{Z}} (\pmb{\theta}) \cdot \pmb{\delta \theta}}{2} \, ,
\end{equation}
where $\pmb{H}_{\ln{Z}}(\pmb{\theta})$ is the Hessian matrix of the function $\ln{Z}$ evaluated in $\pmb{\theta}$ and
\begin{equation}\label{eq:logPhiTaylorExp}
\ln{\frac{\Phi(J+\delta J , \{\tilde{\sigma}\})}{\Phi(J , \{\tilde{\sigma}\})}}  \simeq \frac{\partial \ln{\Phi(J , \{\tilde{\sigma}\})}}{\partial J} \delta J + \frac{1}{2}\frac{\partial ^2 \ln{\Phi(J,\{\tilde{\sigma}\})}}{\partial J^2} \delta J^2 \, .
\end{equation}
In particular, we observe that the first and the second derivatives correspond to expectation values with respect to (w.r.t.) probability distributions that we cannot evaluate explicitly but that we can sample from.
We have that 
\begin{eqnarray}\label{partition_function_derivatives}
(\nabla_{\pmb{w}} \ln{Z(\pmb{\theta})})_j
& = & \mathbb{E}_{\{r\},\{\sigma\}|\pmb{w},J}\left[ \sum_i F_{ij}\sigma_i\right] \, , \nonumber\\
\frac{\partial \ln{Z(\pmb{\theta})}}{\partial J}
& = & \mathbb{E}_{\{r\},\{\sigma\}|\pmb{w},J}\left[\frac{1}{2}\sum_{i\neq j}\sigma_i\sigma_j\Delta(\pmb{r}_i,\pmb{r}_j)\right] \, , \nonumber\\
\end{eqnarray}
where $\mathbb{E}_{\{r\},\{\sigma\}|\pmb{w},J}\left[\cdot\right]$ stands for the expectation value w.r.t. the distribution of the spins and polymer configurations at fixed $\pmb{w}$ and $J$, {\it i.e.} w.r.t. the joint likelihood.
The elements of the Hessian matrix in Eq.~\eqref{eq:logZTaylorExp} read:
\begin{equation}
    \left[ \pmb{H}_{\ln Z}(\pmb{\theta}) \right]=  \left< s_i s_j \right> - \left< s_i \right> \left< s_j \right> = cov(\pmb{s})_{ij} \, ,
\end{equation}
where we use the shorthand notation $\langle \cdot \rangle$ for $\mathbb{E}_{\{r\},\{\sigma\}|\pmb{w},J}[\cdot]$ and the array $\pmb{s}$ is defined as:
\begin{equation*}
\pmb{s} \equiv
\begin{bmatrix}
    \sum_i F_{i1}\sigma_i \\
    \sum_i F_{i2}\sigma_i \\
    \vdots \\
    \vdots \\
    \sum_i F_{if}\sigma_i \\
    \frac{1}{2}\sum_{i\neq j}\sigma_i\sigma_j\Delta(\pmb{r}_i,\pmb{r}_j)
\end{bmatrix}
\in \mathbb{R}^{f+1} \, .
\end{equation*}
%
We developed a simulator for \emph{SEMPER} based on ~\cite{Madras:1988ei, Orlandini1998MMC, GarelOrlandOrlandini1999} that we can employ to obtain samples from the likelihood $P(\{r\}, \{\sigma\}| \pmb{\theta})$ and then use these samples to obtain unbiased estimates of the quantities needed to compute the above Taylor expansions.
Finally, the ratio of the two partition functions in Eq.~\eqref{eq:logZTaylorExp} reads:
\begin{equation}
    \frac{Z(\pmb{\theta}+\pmb{\delta \theta})}{Z(\pmb{\theta})} \simeq \exp\left(\pmb{\delta \theta} \cdot \left<\pmb{s}\right> + \frac{\pmb{\delta \theta}^T \cdot cov(\pmb{s}) \cdot \pmb{\delta \theta}}{2}\right)
\end{equation}
with corrections that are of the order of $\sim\mathcal{O}(\|\pmb{\delta \theta}\|^3)$,
while for the evaluation of the ratio in Eq.~\eqref{eq:logPhiTaylorExp} we use the expressions:
\begin{equation}\label{eq:MeanDLogPhiDJ}
    \frac{\partial \ln{\Phi(J,\{\tilde{\sigma}\})}}{\partial J} = \mathbb{E}_{\{r\}|J,\{\tilde{\sigma}\}}\left[ \frac{1}{2}\sum_{i\neq j}\tilde{\sigma}_i \tilde{\sigma}_j \Delta(\pmb{r}_i,\pmb{r}_j)\right] \, ,
\end{equation}
and
\begin{equation}\label{eq:VarDLogPhiDJ}
    \frac{\partial ^2\ln{\Phi(J,\{\tilde{\sigma}\})}}{\partial J^2} = \mathbb{V}_{\{r\}|J,\{\tilde{\sigma}\}}\left[ \frac{1}{2}\sum_{i\neq j}\tilde{\sigma}_i \tilde{\sigma}_j \Delta(\pmb{r}_i,\pmb{r}_j)\right] \, .
\end{equation}
The expectation~\eqref{eq:MeanDLogPhiDJ} and the variance~\eqref{eq:VarDLogPhiDJ} are computed with respect to the distribution function:
\begin{equation}\label{}
    P(\{r\}|J,\{\tilde{\sigma}\}) = \frac{e^{\frac{J}{2}\sum_{i\neq j}\tilde{\sigma}_i \tilde{\sigma}_j \Delta(\pmb{r}_i,\pmb{r}_j)}}{\Phi(J,\{\tilde{\sigma}\})} \, ,
\end{equation}
which describes the probability for the polymer configurations with spin variables $\{\sigma\}$ having values equal to the observed spin configuration $\{\tilde{\sigma}\}$.
For a more complete derivation and details about the simulator see also Appendix~\ref{app:simulations} and~\ref{app:approximate_mcmc}.

\subsection{Datasets used as inputs for the model and experimental Hi-C data}\label{sec:WhereDatasets}
All the employed datasets were obtained from the ENCODE database~\cite{encode} and through the UCSC genome browser~\cite{ucsc}.

\section{Results: Epigenetic state and spatial conformations of human chromosome 2}\label{sec:Results}

To test the effectiveness of {\it SEMPER} on a case study with real data, we study the spatial conformations of selected portions of human chromosome 2 for different coarse-grain levels of the chromatin fiber.
\subsection{Feature selection and data sets description}\label{sec:FeatureSelection}
As a marker for closed chromatin, we use ChIP-seq tracks of \emph{H3K27me3} for the two biological replicates available for download from the \emph{ENCODE project} database~\cite{encode}. 

As predictors, we use a variety of sequence-derived features and measured binding of transcription factor proteins (see Table~\ref{tab:InputDataDescription} in Appendix~\ref{app:figs} for a complete list of data and their sources).
In particular, we use CpG content as a sequence-derived feature, due to its clear biological relevance in the context of epigenetics. 
It was shown~\cite{Benveniste2014} that TF binding data around transcription start sites (\emph{TSS}) have good predictive power regarding the presence (or the absence) of histone modifications around the same sites.
We therefore employ the same TF binding data used in that paper to build the features matrix $\pmb{F}$ used as input~\eqref{eq:hamiltonian} for determining the external fields $h_i$ of the {\it SEMPER} model.
This will enable a direct comparison between the results of the purely data-driven logistic regression model of~\cite{Benveniste2014} (which, as we noted, represents a limiting case of {\it SEMPER}) and the polymer-based approach, in particular in terms of the interpretation of model results.

\subsection{Predictions of epigenetic states}\label{sec:EpiStatesPrediction}

%
\begin{figure*}
\centering
\includegraphics[width=1.00\linewidth]{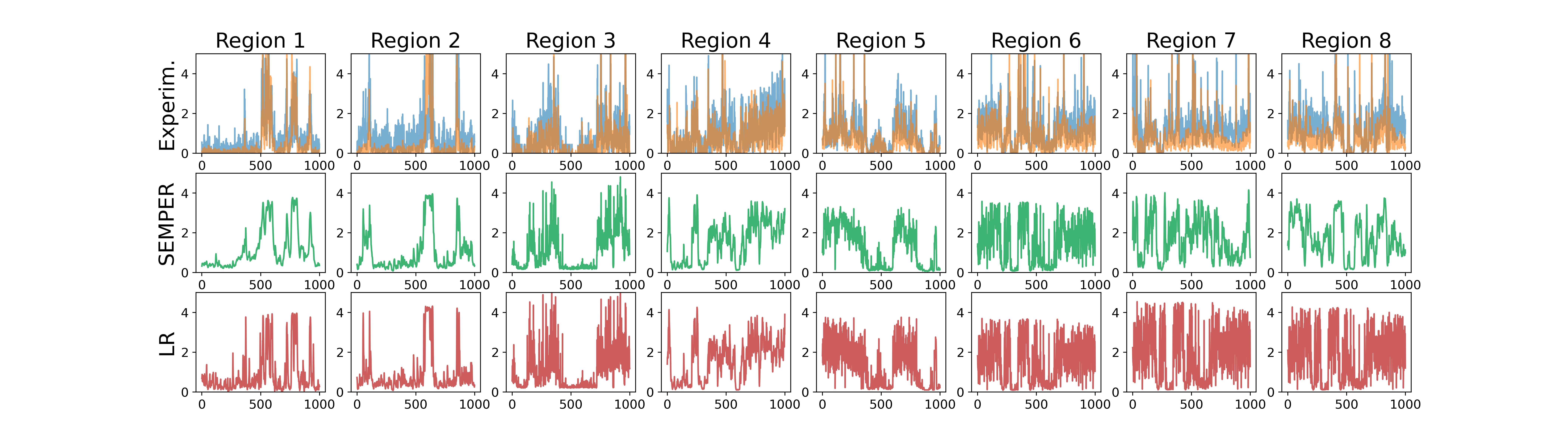}
\caption{
The {\it SEMPER} model allows to recapitulate histone modification patterns.
(Top row)
Comparison between the two biological replicates of the experimentally measured {\it H3K27me3} ChIP-seq tracks~\cite{encode}.
(Middle and bottom row)
Average patterns with respect to the predictive posterior distributions of the spin configurations calculated by, respectively, the {\it SEMPER} and the logistic regression (LR) model.
}
\label{fig:generated_data}
\end{figure*}
\begin{table}[ht]
\centering
\begin{tabular}{ |c|c|c||c|c|c|  }
 \hline
 Start & End & Res. (bps) & Replicates & SEMPER & LR\\
\hline
 218.6 & 219.1 & 0.5k   & $0.894$   & $0.874$    & $0.785$\\
 \hline
 222.0 & 222.5 & 0.5k   & $0.625$   & $0.786$    & $0.752$\\
 \hline
 130 & 131 & 1k   & $0.818$   & $0.621$    & $0.568$\\
 \hline
 111 & 113 & 2k       & $0.794$   & $0.656$    & $0.631$\\
 \hline
 85 & 90 & 5k   & $0.853$   & $0.684$    & $0.612$\\
 \hline
 129 & 134 & 5k       & $0.750$   & $0.584$    & $0.531$\\
 \hline
 95 & 105 & 10k   & $0.694$   & $0.562$    & $0.546$\\
 \hline
 105 & 115 & 10k       & $0.761$   & $0.647$    & $0.638$\\
 \hline
\end{tabular}
\caption{
Mean Pearson correlation coefficients between the predicted epigenetic state patterns and the experimentally measured histone modification ChIP-seq data (two isogenic replicates for each region).
For each region we report also the correlation between the two biological replicates.
We observe the improvement of the predictions of the {\it SEMPER} model compared to logistic regression (LR~\cite{Benveniste2014}) in which the polymeric nature of chromatin is, by construction, not taken into account.
}
\label{tab:pearson_corr_histone}
\end{table}

Having selected a data set and features, we use the approximate inference method we developed to obtain samples from the  intractable posterior distributions for the model parameters.
This then enables us to compute \emph{predictions} of epigenetic state patterns as average simulated configurations of the {\it H3K27me3} mark presence, see Fig.~\ref{fig:generated_data}.

To generate the predictions we use Bayesian model averaging to obtain spin configurations using the posterior predictive distribution:
\begin{equation}\label{eq:PosteriorPredictiveDistr}
    P(\{\sigma\}, \{r\}|\{\tilde{\sigma}\}) = \int d\pmb{\theta} P(\{\sigma\}, \{r\}| \pmb{\theta}) P(\pmb{\theta} | \{\tilde{\sigma}\}) \, .
\end{equation}
This estimator provides also a full account of the uncertainties in the parameters' estimates and the related uncertainties over the predictions.
Table~\ref{tab:pearson_corr_histone} shows the correlations between posterior predicted profiles and the actual ChIP-seq profiles used to train the model, as well as the correlation between the two replicate experimental profiles. Examining the last two columns, we see that the adoption of the magnetic polymer model improves noticeably the predictions if compared to the situation~\cite{Benveniste2014} where the polymeric nature of the chromatin fiber is neglected. The improvement is particularly marked for shorter lengthscales of coarse-graining (500bp, of the same order of magnitude as the length of DNA wrapped around a nucleosome), where SEMPER achieves a correlation level close to the correlation between experimental replicats. 
Since $\sigma_i=1/0$ accounts for the presence/absence of the histone modification, 
portions of the polymer rich in positive spins have the tendency to condense in agreement with the known fact~\cite{Spakowitz2018} that {\it H3K27me3} is associated with transcriptional repression {\it via} the formation of heterochromatic regions.

\subsection{Influence of polymeric terms in estimating the parameters}\label{sec:PolymerTermsAndParameters}
\begin{figure}[ht]
    \centering
    \includegraphics[width=0.98\linewidth]{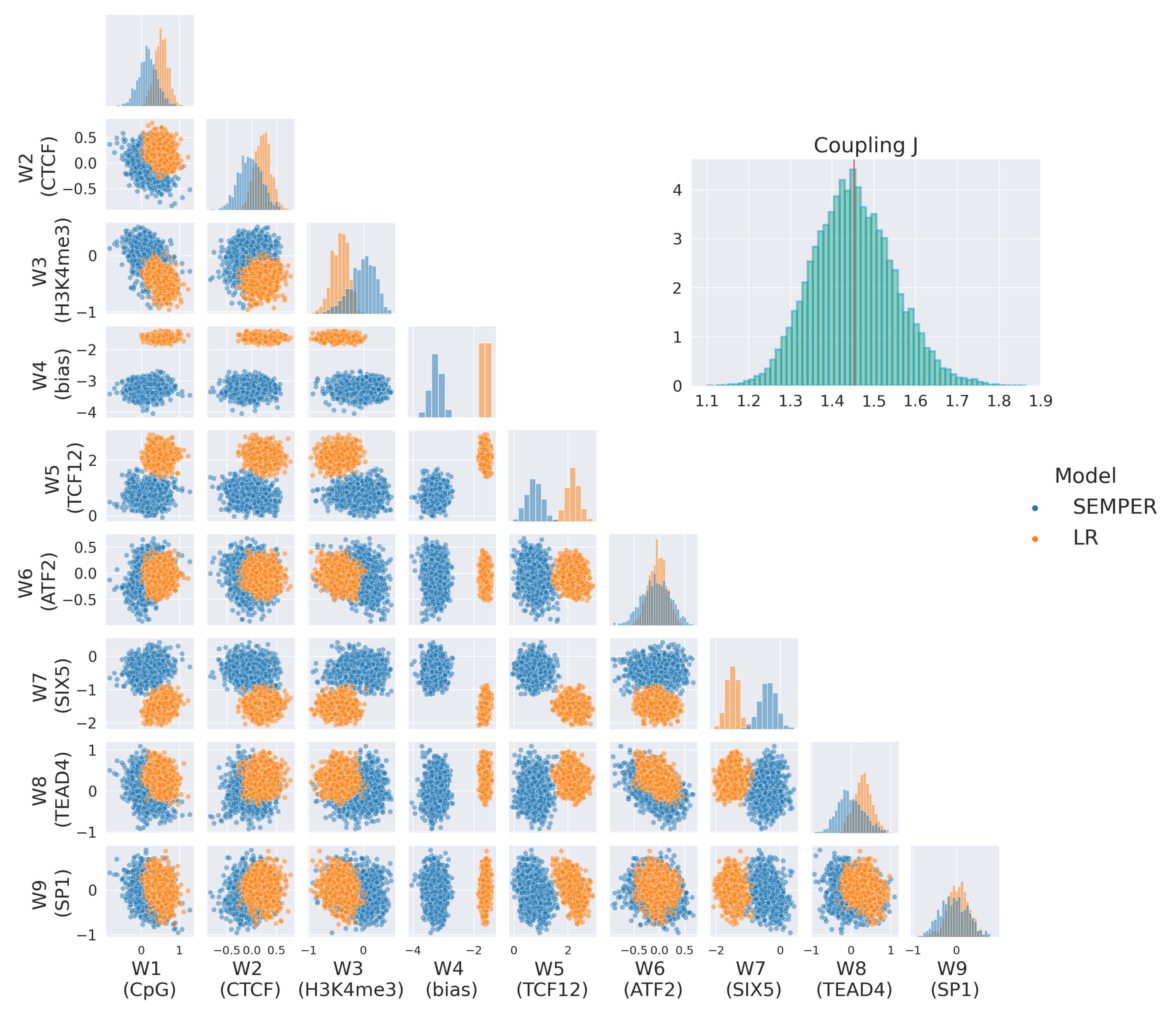}
    \caption{
    {\it SEMPER} and the approximate Monte Carlo scheme developed in this work allow us to obtain samples from the otherwise intractable posterior distribution of the model.
    Here, posterior distributions for the region \emph{chr2:218,600,000-219,100,000} are shown.
    Notice the quantitative differences with the logistic regression (LR) posterior distributions.
    }
    \label{fig:qamhi_posterior}
\end{figure}

Our results show quantitatively
that the incorporation of a polymer structure leads to improved recapitulation of chromatin state.
Even more importantly Fig.~\ref{fig:qamhi_posterior} also suggests a major difference in terms of the possible interpretation of the results obtained.
The panels show samples from joint posteriors over pairs of weights obtained from our approximate Monte Carlo 
algorithm (blue dots) and from the logistic regression (LR) model previously adopted in~\cite{Benveniste2014}.
The comparison illustrates a striking difference between the estimates: first of all, the estimate of the interaction parameter $J$ is significantly different from zero, indicating that the data strongly favors an explanation which includes the polymer nature of the chromatin.
{\it SEMPER} estimates of weights tend to have larger uncertainty and to be generally of smaller average magnitude compared to the LR estimates.
Intriguingly, {\it SEMPER} infers a very small weight for the feature corresponding to CpG density, consistent with the alternative chromatin regulating mechanisms underpinning CpG methylation and polycomb-repressor complex-mediated deposition of {\it H3K27me3}~\cite{reddington2013redistribution}.
In turn, discrepancies in weight estimates can lead to very significant differences in terms of interpretation: our results suggest that the importance of TFs in shaping the chromatin landscape might have been overestimated by purely data-driven efforts such as~\cite{Benveniste2014}, while many conformational features can largely emerge as a result of polymer mechanisms amplifying relatively subtle genetic signals.

%

\subsection{Bottom-up modeling leads to realistic interaction patterns}\label{sec:BottomUpAndInteractions}

\begin{figure*}
        \centering
        \includegraphics[width=.32\textwidth]{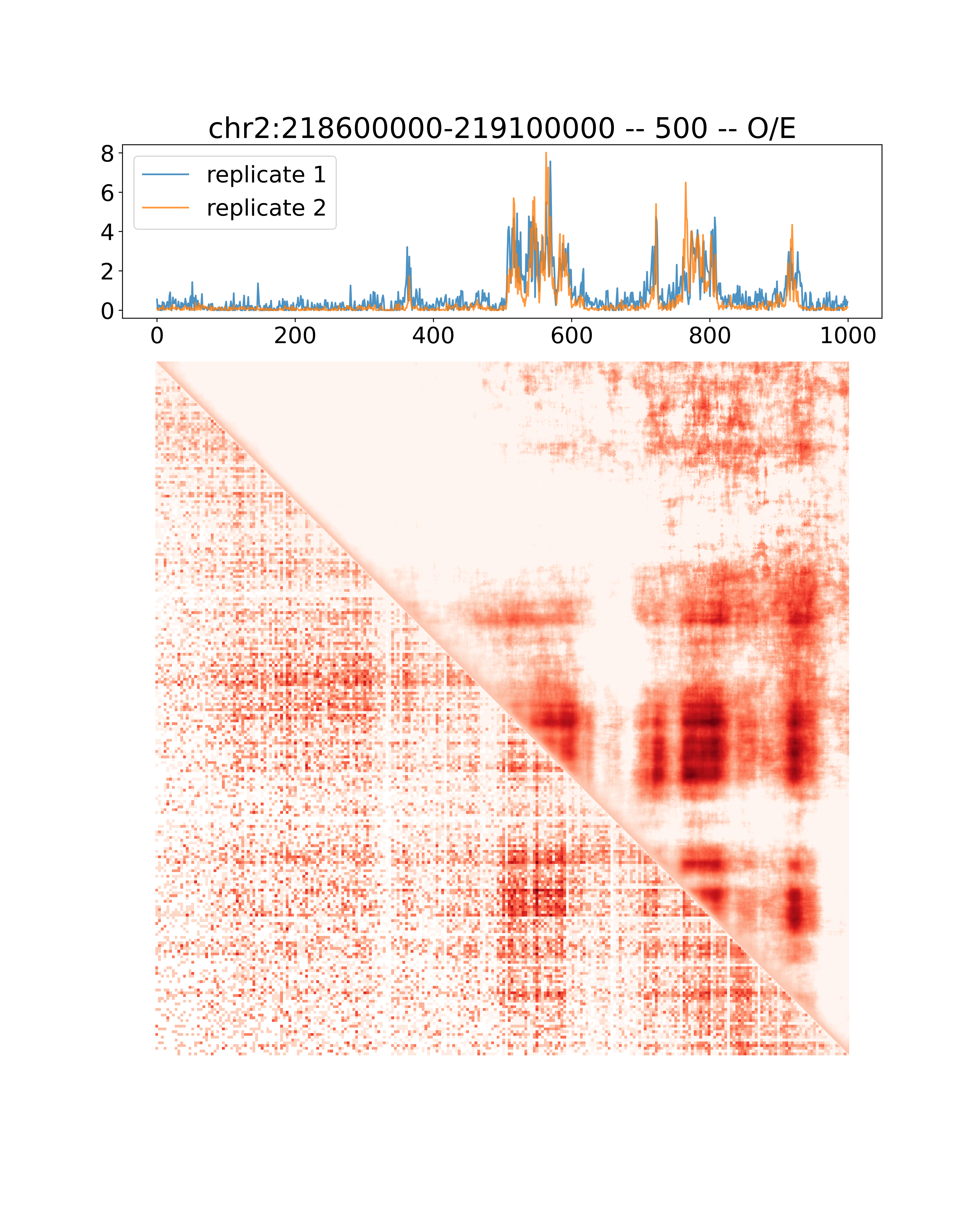}
        \includegraphics[width=.32\textwidth]{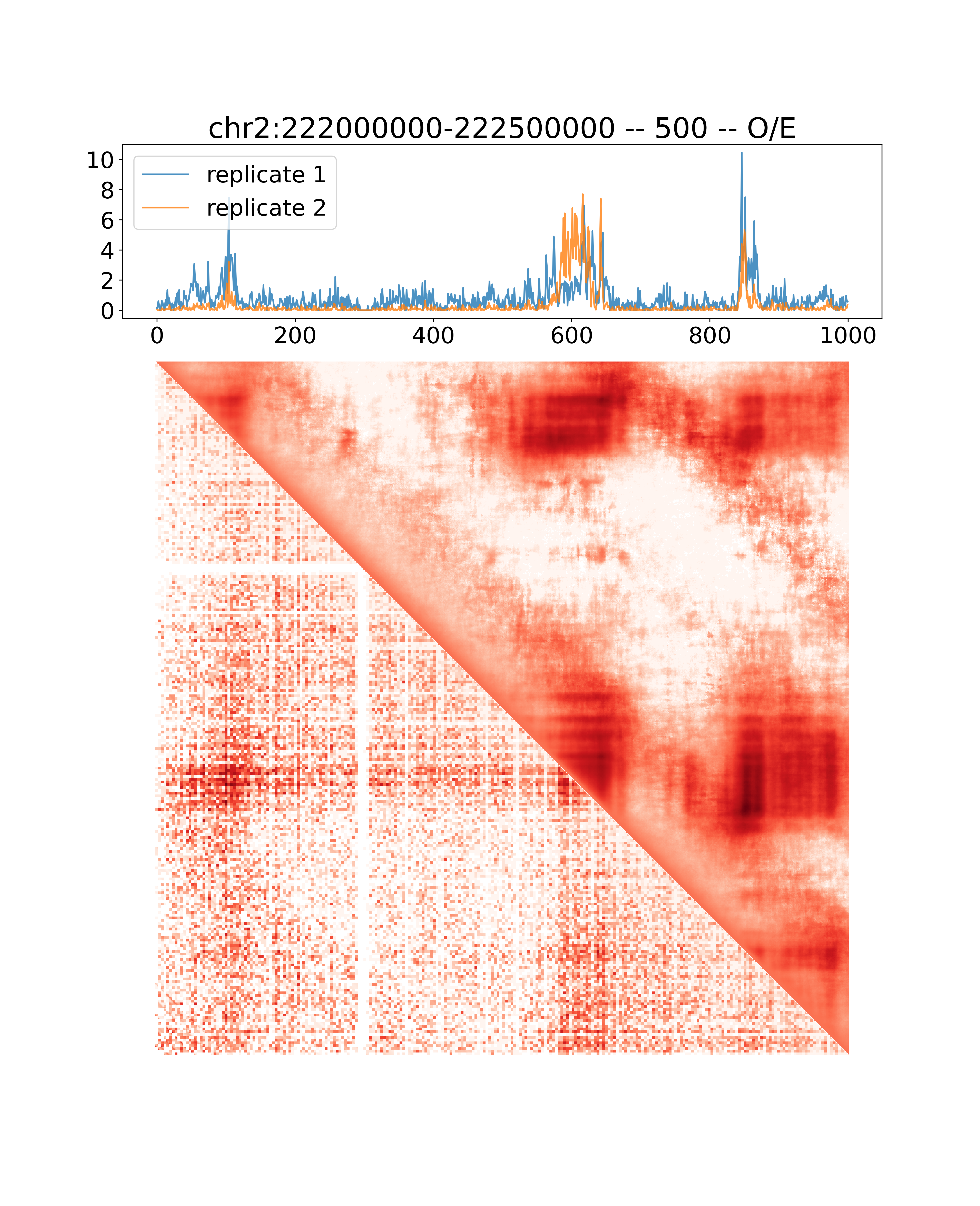}
        \includegraphics[width=.32\textwidth]{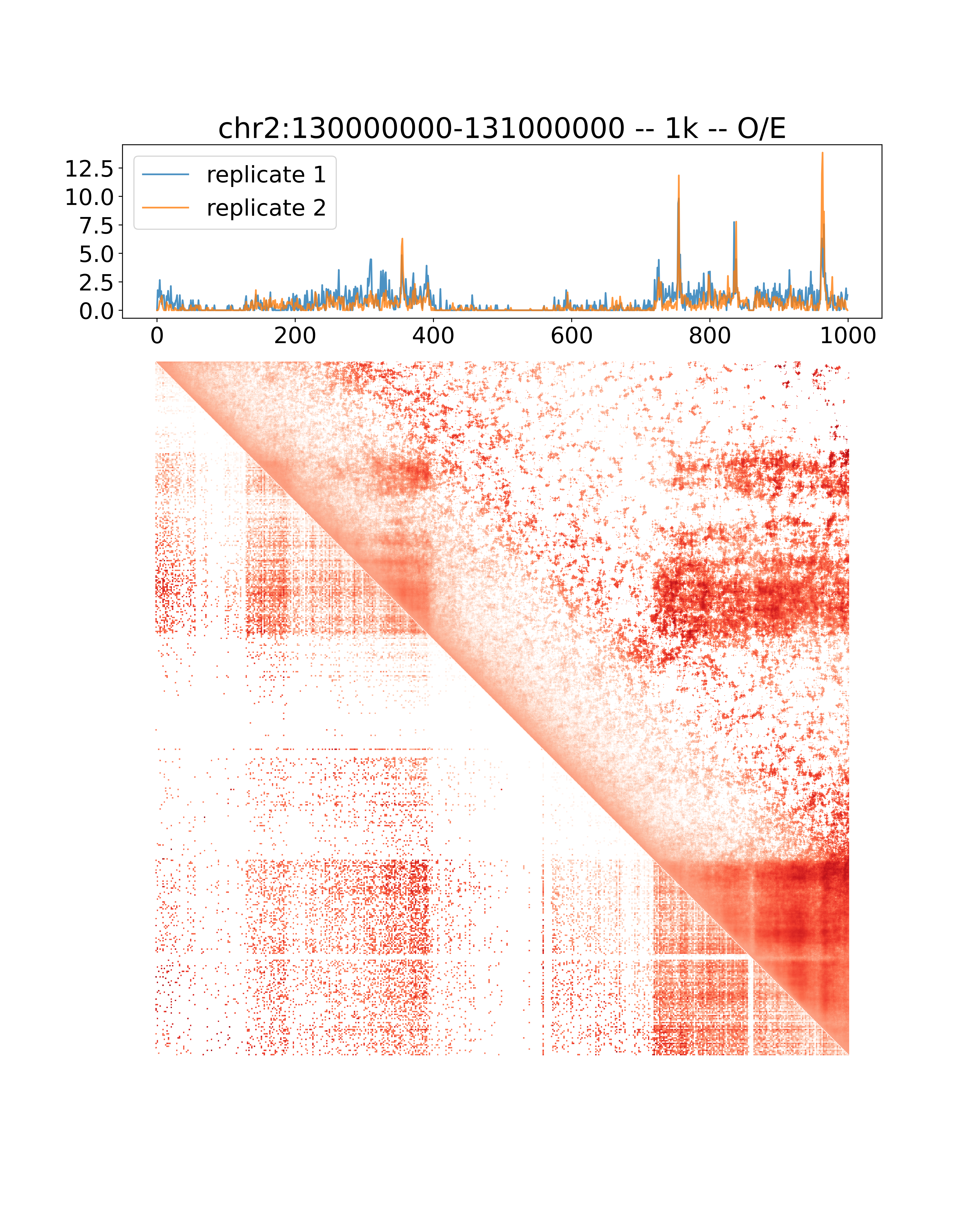}
        \includegraphics[width=.32\textwidth]{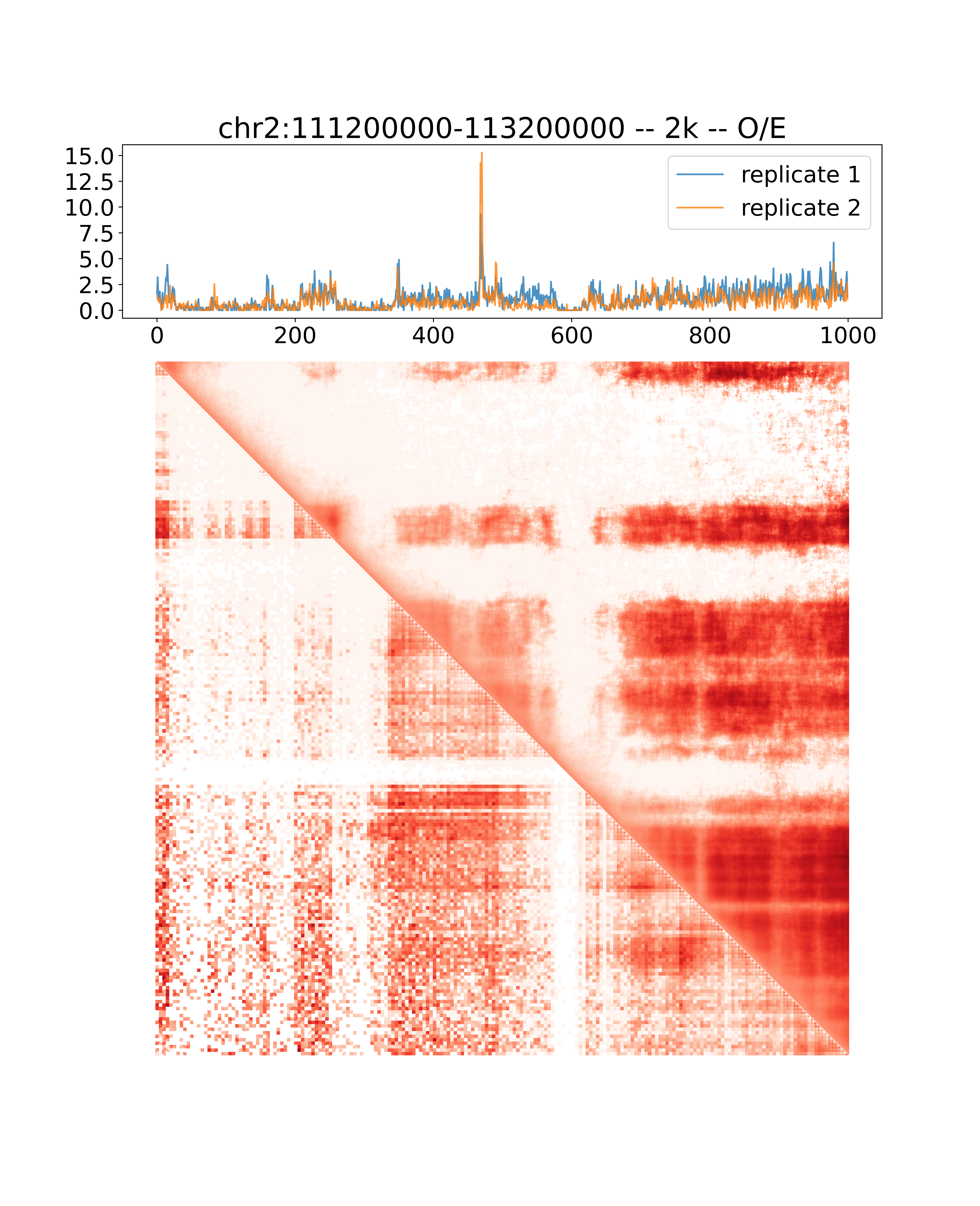}
        \includegraphics[width=.32\textwidth]{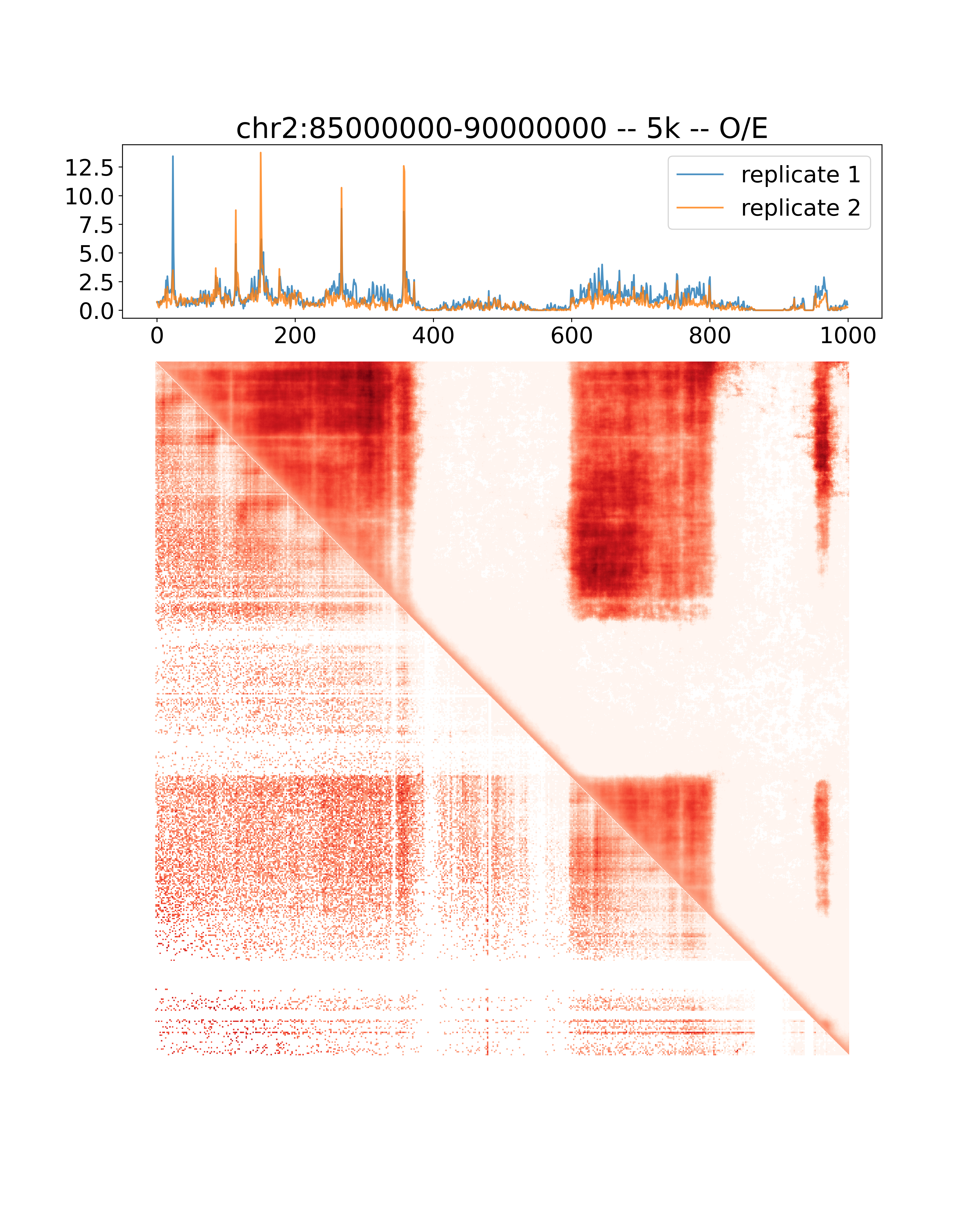}
        \includegraphics[width=.32\textwidth]{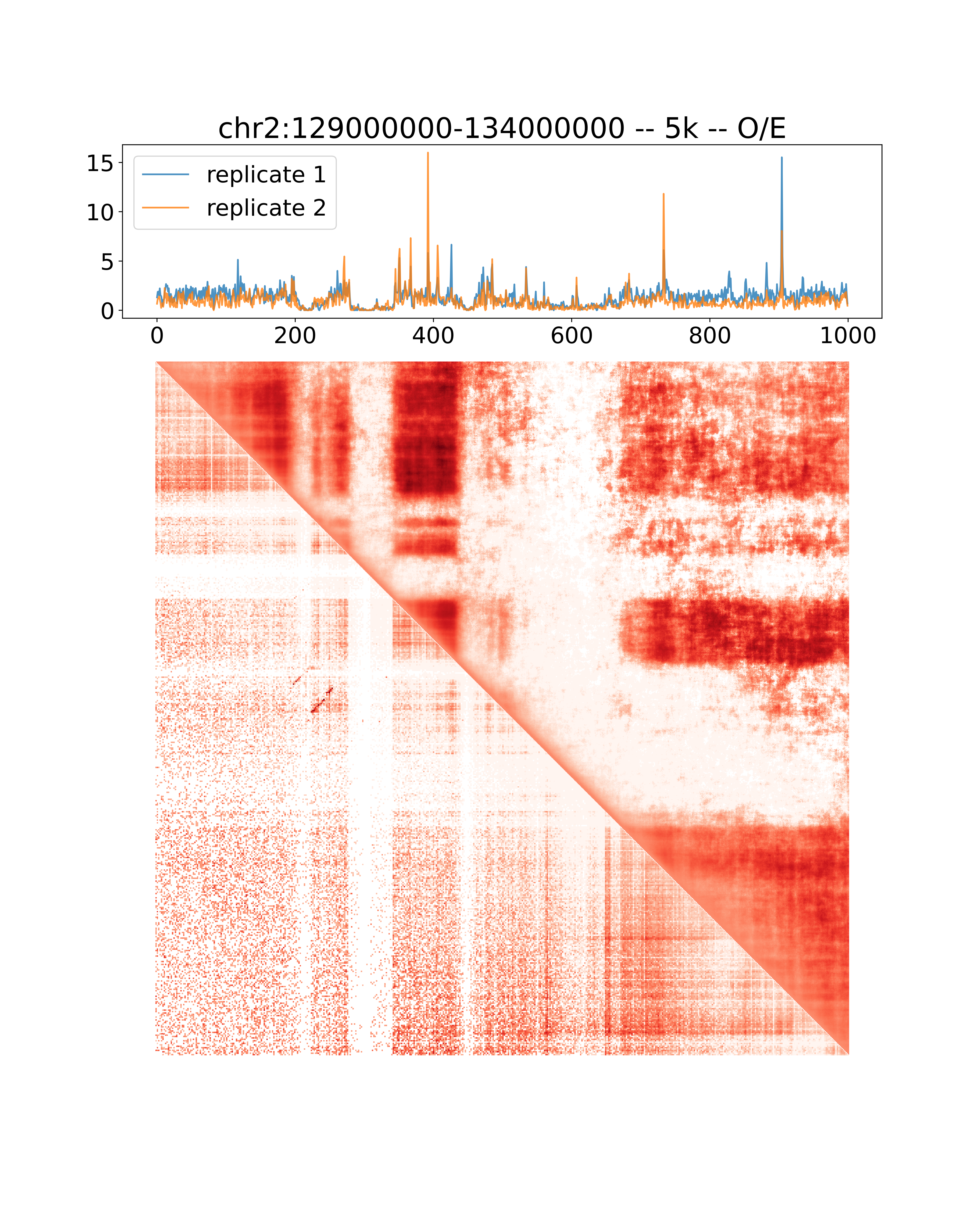}
        \includegraphics[width=.32\textwidth]{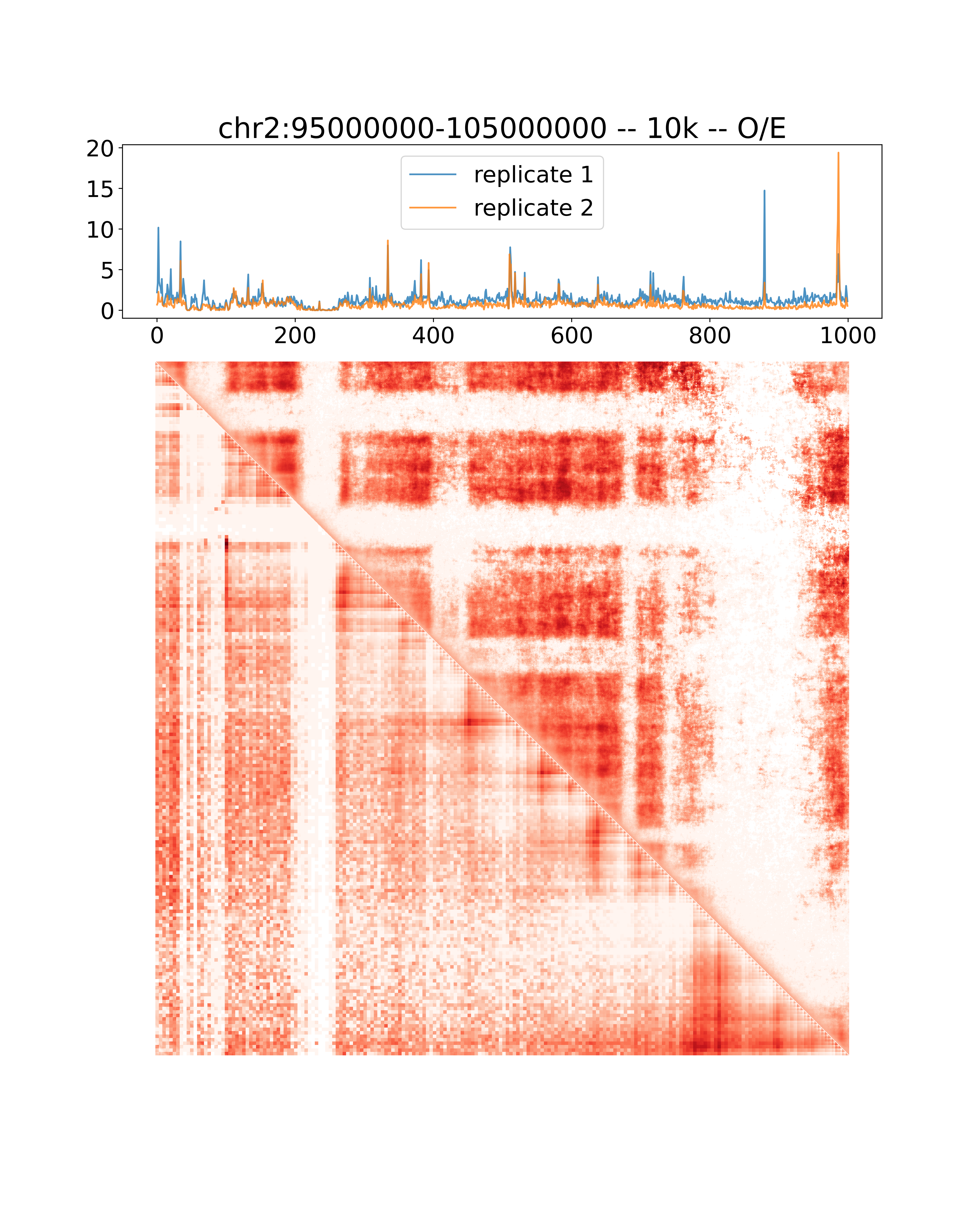}
        \includegraphics[width=.32\textwidth]{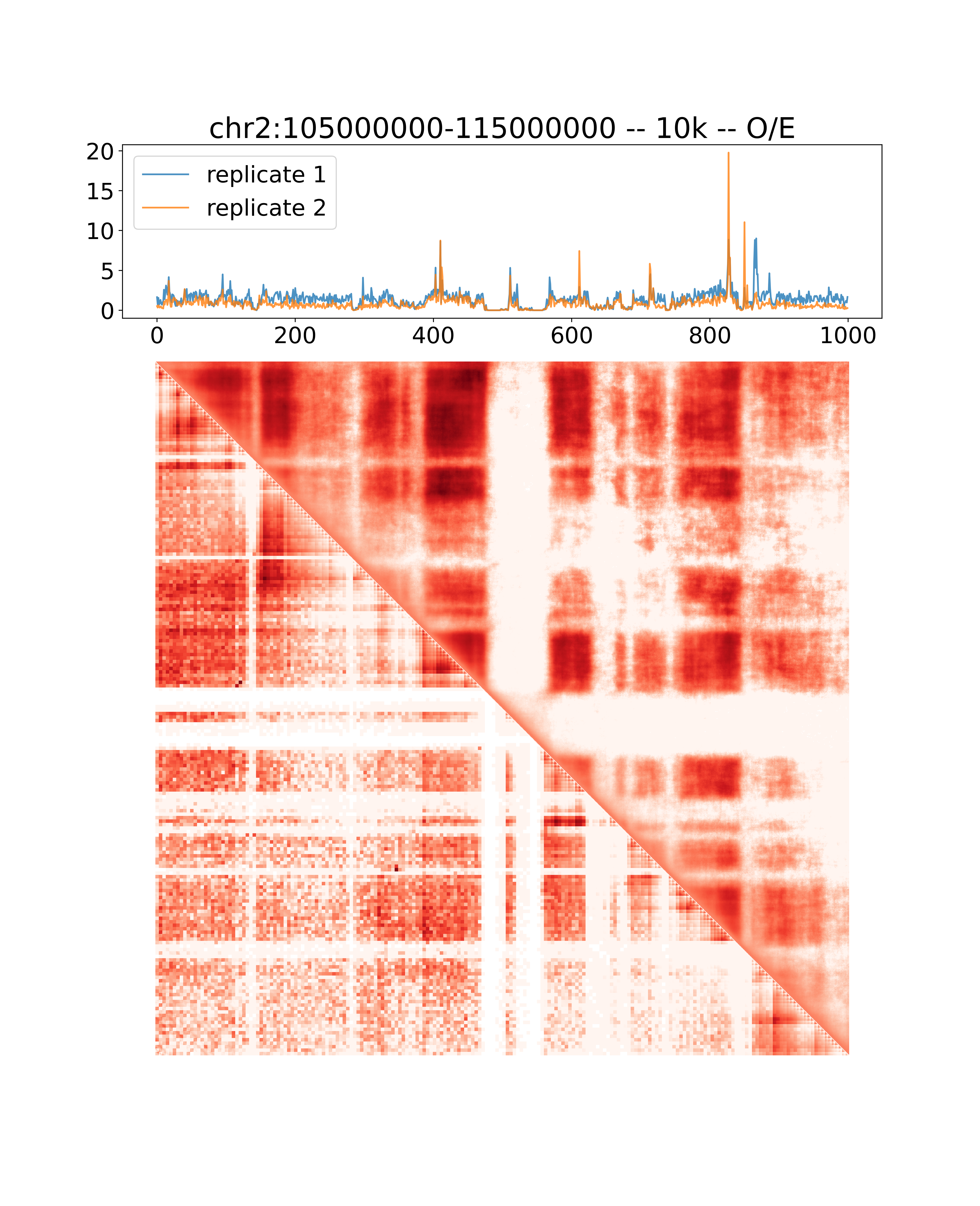}            
        \caption{
        Comparison of experimental and calculated ChIP-seq tracks and Hi-C matrices.
        In each panel:
        (i) the top plot shows the {\it H3K27me3} ChIP-seq tracks for the two employed biological replicates;
        (ii) the bottom plot shows the comparison between the experimentally measured Hi-C maps (lower triangular part) and the calculated ones from our model chromatin conformations (upper triangular part).
        The intensity of the pixels corresponds to the normalized log-observed-{\it vs.}-expected number of reads.
        }
        \label{fig:contact_maps}
\end{figure*}

Besides being able to predict epigenetic state patterns from sequence and TF binding data alone, the model provides us with realistic putative chromatin conformations. 
To demonstrate their plausibility, we compute the corresponding contact maps between chromatin loci and compare those to the ones measured through Hi-C technologies~\cite{Dekker2009,rao20143d}.
Hi-C, which is typically based on a suitable combination of cross-linking, ligation and sequencing of nuclear DNA, constitutes a powerful and worldwide established protocol for the systematic detection of chromatin-chromatin interactions.
The result of the experiment can be intuitively displayed in the form of {\it contact maps}, whose entries correspond to the number of detected pairwise contacts between the two generic fragments $i$ and $j$ along the genomic sequence of interest.

In order to compare with experiments, contact matrices from simulations are computed in the 
following way.
For a given chromatin conformation, a contact between monomers $i$ and $j$ is counted if
$||\pmb{r}_j - \pmb{r}_i||^2\leq r_c^2$
where $r_c$ is a suitable cutoff contact distance.
Different values of $r_c^2$ in the range $[5d^2-20d^2]$ where $d$ is the employed spatial discretization of the chromatin fiber (Sec.~\ref{sec:ModelHamiltonian}), were used without displaying substantial differences in the emerging patterns in the contact maps.
The experimental {\it vs.} the simulated matrices are shown in Fig.~\ref{fig:contact_maps}.
Importantly, like the authors in~\cite{Spakowitz2018}, we never include any structural information about the chromatin organization itself in the model.
Our aim is to explore to what extent a bottom-up approach integrating sequence and polymer models can recapitulate the contact maps of the chromatin, rather than obtaining the highest possible predictive accuracy.

From a detailed inspection of Fig.~\ref{fig:contact_maps}, it is clear that not all the details of the experimental maps can be captured from our model, and some spurious contacts are easily discernible at the boundaries of the simulated matrices (likely the result of working with a finite polymer chain, and not with a full chromosome as in~\cite{RosaPlos2008}). 
Despite these artifacts, we still manage to obtain realistic patterns of interaction consistent with experimental results.
This is particularly remarkable considering that we have used data for just a single kind of histone modifications and only a few transcription factors, which may not be expected to recapitulate much information about the geometry of the chromatin filament. 

These results suggest a central role for the polymer nature of the chromatin in the spatial organization of the genome, integrating post-transcriptional modifications of the histone tails and binding of transcription factor proteins. 
Furthermore, these results suggest a potential usage of the model and methods presented in this work to investigate the relationship between the various existing histone modifications and different structural features of the spatial genomic organization at different length scales.

\section{Discussion and conclusions}\label{sec:DiscConcls}
In this work we have introduced {\it SEMPER}, a statistical model for integrating epigenomic data (such as histone marks and TF binding) in a simple and intuitive polymer model of the chromatin fiber.
{\it SEMPER} aims to establish a conceptual bridge between two lines of research in epigenomic data modelling which have been quite separate up to now. 
{\it SEMPER} can be viewed as an extension of the (simpler) logistic regression (LR) predictor~\cite{Benveniste2014}, and hence connects to the rich line of research which develops machine learning methods for epigenomic prediction and imputation.
We show that {\it SEMPER} improves on LR on a number of example regions, likely due to the introduction physically motivated local correlations between epigenomic marks at different monomers on a self-interacting chromatin filament.
On the other hand, {\it SEMPER} is a {\it bona fide} statistical polymer model and hence fully mechanistic. 

Remarkably, we show that {\it SEMPER} is able to predict non-trivial structural features  at a very high resolution without using any structural data and without including any {\it a priori} information or mechanistic knowledge concerning how chromatin folding works. 

Our work confirms and strengthens previous results such as~\cite{Nicodemi2018,DiPierro2016,DiPierro2017} on the importance of the polymeric nature of chromatin, while providing a new framework to integrate seamlessly the (epi)genomic data directly within a polymer model in a transparent and interpretable way.

Despite its conceptual simplicity, Bayesian inference in {\it SEMPER} is highly intractable, preventing the deployment of standard sampling algorithms.
A major contribution of our work has been the development of a novel approximate inference method that allows parameter estimation and uncertainty quantification.
This algorithm is in principle applicable to a broad class of models where estimation of the partition function relies on computation of moments using model simulations. 

While we believe {\it SEMPER} constitutes an important conceptual step, several improvements and future directions are possible.
From the biological point of view, our work focussed on modelling a single histone mark, {\it H3K27me3}.
It was used because of its well-known ubiquitous role in the formation of heterochromatic regions; nonetheless, a possible future development should be to consider including multiple spin variables for each monomer in order to account for the many different histone marks shaping the architecture of the chromatin filament.
From the modelling point of view, we chose to introduce the effect of sequence features via a linear external field.
This is natural as a first step and because of its connection with the logistic regression framework, but it can certainly be revisited to include more complex nonlinear effects.
From the computational point of view, the cost of the simulations based on the pivot algorithm~\cite{Madras:1988ei} remains high, preventing for the time being accurate fine-grain (and sufficiently large-scale!) treatment of the whole chromosomes. 
This drawback is maximally due to the need of performing extensive polymer simulations for each step of the posterior sampling algorithm. 
Future work devoted to finding more efficient sampling procedures by utilizing concepts from, {\it e.g.}, advanced machine learning~\cite{cranmer2020frontier} will likely further push the boundaries of physics-based models to analyze epigenomic data.

\begin{acknowledgments}
AR acknowledges networking support by the COST Action CA18127 (INC).
\end{acknowledgments}

\clearpage

\onecolumngrid

\appendix

\section{Simulation details}\label{app:simulations}

\subsection*{A hybrid pivot algorithm}
We describe the chromatin filament and the experimentally observed epigenetic pattern in terms of a {\it magnetic polymer model}~\cite{GarelOrlandOrlandini1999}.
The polymeric nature of the chromatin fiber is modeled as a self-avoiding walk (SAW) on the $3D$ simple cubic lattice~\cite{Madras:1988ei}.
The SAW is made of $N$ monomers linearly connected by $N-1$ bonds of length equal to one lattice unit ,
and each monomer $i = 1, ..., N$ bears a binary Ising-like spin variable $\sigma_i \in \{0,1\}$.
Then, the Hamiltonian ${\mathcal H}$ governing the physical behavior of the model chromatin filament is given by the expression:
\begin{eqnarray}\label{eq:hamiltonian_app}
{\mathcal H}(\{r\}, \{\sigma\} | \pmb{\theta}) 
& = & {\mathcal H}_{\rm SAW}(\{r\})\, -\,\frac{J}{2}\sum_{i\neq j}\sigma_{i}\sigma_{j}\Delta(\pmb{r}_i,\pmb{r}_j) -\sum_{i}h_i(\pmb{w})\sigma_i \nonumber\\
& \equiv & {\mathcal H}_{\rm SAW}(\{r\}) + {\mathcal H}_{\rm Ising}(\{r\}, \{\sigma\} | J) + {\mathcal H}_{\rm ext}(\{\sigma\} | \pmb{w}) \, ,
\end{eqnarray}
where the compact notations
$\{r\} \equiv ({\mathbf r}_1, ..., {\mathbf r}_N)$
and
$\{\sigma\} \equiv (\sigma_1, ..., \sigma_N)$
are for the whole sets of monomers' spatial coordinates and the corresponding Ising-like spin variables.
The meaning of the various terms in~\eqref{eq:hamiltonian} is the following:
\begin{itemize}
    \item
    $\mathcal{H}_{\rm SAW}(\{r\})$ is the pure polymer contribution.
    It is $= +\infty$ if the polymer is self-crossing ({\it i.e.}, non self-avoiding) and zero otherwise~\cite{Madras:1988ei}.
    \item
    $J>0$ is the coupling constant between two spins.
    The spin-spin interaction is limited to pairs of monomers that are nearest-neighbors on the lattice, {i.e.}
    $\Delta(\mathbf{r}_i,\mathbf{r}_j)$ takes value $=+1$ if $|\mathbf{r}_i-\mathbf{r}_j| = d$ and $=0$ otherwise.
    \item
    $h_i, \, i=1,...,N$ are the local external fields.
    In our model, their values are expressed by linear combinations of $f$ selected sequence and TF-binding ``features''. One of the features is a constant throughout monomers and we call it ``bias'' and changing the weight associated to it allows to modulate the overall external fields' magnitude.
    $$
    h_i(\pmb{w}) = \sum_{\alpha=1}^f F_{i\alpha}w_{\alpha} \, ,
    $$
    where $\pmb{F}$ is the $N\times f$ matrix containing the features' values for all monomers' positions and $\pmb{w}$ is the vector  ``weights'' that assign different importance to the different experimentally observed features.
    
    \item 
    $\theta \equiv ( J, \pmb{w})$ is a short-hand notation for all the free parameters featured in the model Hamiltonian~\ref{eq:hamiltonian_app}.
\end{itemize}

Magnetic polymer conformations $(\{r\}, \{\sigma\})$ at fixed values of the parameters $\pmb{\theta} = (J, w_1, ..., w_f)$ are obtained by sampling from the canonical distribution
$$
P(\{r\}, \{\sigma\} | \pmb{\theta}) \sim e^{-{\mathcal H}(\{r\}, \{\sigma\} | \pmb{\theta}) / \kappa_B T} \, .
$$
In this work, the sampling is performed with a standard Metropolis-Hastings~\cite{Metropolis1953} Monte Carlo algorithm.
Similarly to~\cite{GarelOrlandOrlandini1999,MichielettoColiPRL2019} we perform separate moves for the polymer's and spins' degrees of freedom.
Each polymeric move combines one global \emph{pivot} move and
$\sim \mathcal{O}(N)$ local moves, see~\cite{Madras:1988ei}.
If the resulting polymer is not self-avoiding the move is rejected, otherwise it is accepted with the standard Metropolis-Hasting probability.
For the spins degrees of freedom we perform $\sim \mathcal{O}(N)$ single spin flips.

\subsection*{Multiple Markov Chains}
In the high-$J$ ({\it i.e.}, low-temperature) regime, the simple Monte Carlo algorithm gets unavoidably stuck into low-energy metastable polymer conformations resulting in highly inefficient sampling.

To solve this issue, we resort to the Multiple Markov Chains (MMC) method~\cite{Orlandini1998MMC}. 
The basic idea of MMC is to run simultaneously a certain number $p$ of simulations at different temperatures $T_1 > T_2> ... > T_p$.
Then, every once in a while, a pair of conformations from the ensembles at temperatures $T_i$ and $T_{i+1}$ are randomly picked and swapped with each other with a certain probability.
In more detail, by denoting by $\pi_{\mathcal{S}}(T)$ the Boltzmann weight of the state $\mathcal{S}$ at temperature $T$, the swap between states ``$i$'' and ``$i+1$'' is taken with probability: 
\begin{equation}
    \alpha_{\rm swap}(i, i+1) = \min \left( 1,\,\frac{\pi_{\mathcal{S}_{i+1}}(T_i) \, \pi_{\mathcal{S}_i}(T_{i+1})}{\pi_{\mathcal{S}_i}(T_i) \, \pi_{\mathcal{S}_{i+1}}(T_{i+1})} \right) \, .
\end{equation}
Now, if we call $\Gamma$ the space of all possible states of the magnetic polymer, it can be shown that the procedure described above constitutes a Markov chain in $\Gamma^p$ with joint stationary distribution given by the expression:
\begin{equation}\label{eq:MMCStationaryDistr}
    P(\mathcal{S}_{1},\mathcal{S}_{2},\ldots ,\mathcal{S}_{p})\propto \prod_{i=1}^p\,\pi_{\mathcal{S}_i}(T_i) \, .
\end{equation}
The factorization in~\eqref{eq:MMCStationaryDistr} allows us to say that for each individual chain $i$ the sequence of states is asymptotically distributed according to the canonical distribution with the Hamiltonian~\eqref{eq:hamiltonian_app} and temperature

\section{Approximate inference method}\label{app:approximate_mcmc}

\subsection*{Approximate method for Metropolis-Hastings sampling}
As mentioned in the main text the posterior distribution of the problem at hand is not tractable and reads:
\begin{equation}
    P(\pmb{\theta}|\{\tilde{\sigma}\}) \propto P(\{\tilde{\sigma}\}|\pmb{\theta}) P(\pmb{\theta}) = \frac{e^{-{\mathcal H}_{\rm ext}(\{\tilde{\sigma}\} | \pmb{w})}}{Z(\pmb{\theta})}\Phi(J,\{\tilde{\sigma}\})P(\pmb{\theta}) \, .
\end{equation}
If one attempts to sample from this probability distribution with a Metropolis-Hastings MCMC algorithm~\cite{Metropolis1953}, then the acceptance ratio is
\begin{equation}\label{eq:MHAccRatio}
    \alpha_{\rm MC} = \min\left( 1,\, \frac{e^{-\mathcal{H}_{\rm ext}(\{\tilde{\sigma}\} | \pmb{w} + \pmb{\delta w})} p(\pmb{\theta} + \pmb{\delta \theta})}{e^{-\mathcal{H}_{\rm ext}(\{\tilde{\sigma}\} | \pmb{w})} p(\pmb{\theta})}\cdot\frac{Z(\pmb{\theta})}{Z(\pmb{\theta} + \pmb{\delta \theta})} \cdot \frac{\Phi(J + \delta J,\{\tilde{\sigma}\})}{\Phi(J,\{\tilde{\sigma}\})}\right) \, ,
\end{equation}
where $\pmb{\theta} \equiv (J,\pmb{w})$ is the current state of the Markov chain, while $\pmb{\theta} + \pmb{\delta \theta} \equiv(J + \delta J,\pmb{w} + \pmb{\delta w})$ is the proposed new state.
In order to compute the ratio~\eqref{eq:MHAccRatio}, one needs to evaluate the two ratios
$\frac{Z(\pmb{\theta}+\pmb{\delta \theta})}{Z(\pmb{\theta})}$
and
$\frac{\Phi(J+\delta J,\{\tilde{\sigma}\})}{\Phi(J,\{\tilde{\sigma}\})}$
between intractable normalization constants.

The method we propose here is based on the fact that one can approximate the above ratios by writing the following Taylor expansions:
\begin{equation}
\ln{\frac{Z(\pmb{\theta}+\pmb{\delta \theta})}{Z(\pmb{\theta})}}
= \pmb{\delta \theta}\cdot\nabla_{\pmb{\theta}} \ln{Z(\pmb{\theta})} + \frac{\pmb{\delta \theta}^T \cdot \pmb{H}_{\ln{Z}} (\pmb{\theta}) \cdot \pmb{\delta \theta}}{2} + \mathcal{O}(\|\pmb{\delta \theta}\|^3) \, 
\end{equation}
where $\pmb{H}_{\ln{Z}}(\pmb{\theta})$ is the Hessian matrix of the function $\ln{Z}$ evaluated in $\pmb{\theta}$ and
\begin{equation}
    \ln{\frac{\Phi(J+\delta J,\{\tilde{\sigma}\})}{\Phi(J,\{\tilde{\sigma}\})}} = \frac{\partial \ln{\Phi(J,\{\tilde{\sigma}\})}}{\partial J} \delta J + \frac{1}{2}\frac{\partial ^2 \ln{\Phi(J,\{\tilde{\sigma}\})}}{\partial J^2} \delta J^2 + \mathcal{O}(\delta J ^3) \, .
\end{equation}
Then, by defining the array
\begin{equation*}
\pmb{s} \equiv
\begin{bmatrix}
    \sum_i F_{i1}\sigma_i \\
    \sum_i F_{i2}\sigma_i \\
    \vdots \\
    \vdots \\
    \sum_i F_{if}\sigma_i \\
    \frac{1}{2}\sum_{i\neq j}\sigma_i\sigma_j\Delta(\pmb{r}_i,\pmb{r}_j)
\end{bmatrix}
\in \mathbb{R}^{f+1} \, ,
\end{equation*}
we have that 
\begin{eqnarray}\label{partition_function_derivatives-Appendix}
(\nabla_{\pmb{w}} \ln{Z(\pmb{\theta})})_j
& = & \mathbb{E}_{\{r\},\{\sigma\}|\pmb{w},J}\left[ \sum_i F_{ij}\sigma_i\right] = \,\mathbb{E}_{\{r\},\{\sigma\}|\pmb{w},J}\left[ s_j\right]\, , \, \text{for} \, j=1\ldots f \nonumber\\
\frac{\partial \ln{Z(\pmb{\theta})}}{\partial J}
& = & \mathbb{E}_{\{r\},\{\sigma\}|\pmb{w},J}\left[\frac{1}{2}\sum_{i\neq j}\sigma_i\sigma_j\Delta(\pmb{r}_i,\pmb{r}_j)\right] =  \,\mathbb{E}_{\{r\},\{\sigma\}|\pmb{w},J}\left[ s_{f+1}\right] \, ,
\end{eqnarray}
where
$\mathbb{E}_{\{r\},\{\sigma\}|\pmb{w},J}\left[\cdot\right]$
stands for the expectation value of the corresponding quantity w.r.t. the distribution function of the spins and polymer configurations at fixed $\pmb{w}$ and $J$, {\it i.e.} w.r.t. the joint likelihood.
Finally, the elements of the Hessian matrix read:
\begin{eqnarray}\label{eq:Hessian_matrix_elements}
\left[ \pmb{H}_{\ln Z}(\pmb{\theta}) \right]_{ij}
& = & \frac{\partial}{\partial \theta_i \partial \theta_j} \ln Z(\pmb{\theta}) =  \frac{\partial}{\partial \theta_i}\left[ \frac{1}{Z(\pmb{\theta})} \sum_{\{r\}, \{\sigma\}} s_j e^{-\mathcal{H}(\{r\}, \{\sigma\} | \pmb{\theta})}\right] \nonumber\\
& = & \frac{1}{Z(\pmb{\theta})}\sum_{\{r\}, \{\sigma\}} (s_j s_i) e^{-\mathcal{H}(\{r\}, \{\sigma\} | \pmb{\theta})} - \frac{1}{Z^2(\pmb{\theta})} \sum_{\{r\}, \{\sigma\}} s_j e^{-\mathcal{H}(\{r\}, \{\sigma\} | \pmb{\theta})} \sum_{\{r\}, \{\sigma\}} s_i e^{-\mathcal{H}(\{r\}, \{\sigma\} | \pmb{\theta})} \nonumber\\
& = & \left< s_i s_j \right> - \left< s_i \right> \left< s_j \right> = cov(\pmb{s})_{ij} \, ,
\end{eqnarray}
where we use the shorthand notation $\langle \cdot \rangle$ for $\mathbb{E}_{\{r\},\{\sigma\}|\pmb{w},J}[\cdot]$. 

Using our simulator one can obtain samples from the likelihood
$P(\{r\}, \{\sigma\}| \pmb{\theta})$
and then use these samples to obtain unbiased estimates of the quantities needed to compute the above Taylor expansion.
Indeed, the first ratio reads:
\begin{equation}
    \frac{Z(\pmb{\theta}+\pmb{\delta \theta})}{Z(\pmb{\theta})} = \exp\left(\pmb{\delta \theta} \cdot \left<\pmb{s}\right> + \frac{\pmb{\delta \theta}^T \cdot cov(\pmb{s}) \cdot \pmb{\delta \theta}}{2}\right)(1 + \mathcal{O}(\|\pmb{\delta \theta}\|^3)) \, .
\end{equation}
To evaluate the second Taylor expansion we notice that:
\begin{equation}
    \frac{\partial \ln{\Phi(J,\{\tilde{\sigma}\})}}{\partial J} = \frac{1}{\Phi(J,\{\tilde{\sigma}\})}\sum_{\{r\} \in\, SAW}\frac{1}{2}\sum_{i\neq j}\tilde{\sigma}_i \tilde{\sigma}_j \Delta(\pmb{r}_i,\pmb{r}_j) e^{\frac{J}{2}\sum_{i\neq j}\tilde{\sigma}_i \tilde{\sigma}_j \Delta(\pmb{r}_i,\pmb{r}_j)} \, ,
\end{equation}
corresponding to the expectation value of
$\frac{1}{2}\sum_{i\neq j}\sigma_i \sigma_j \Delta(\pmb{r}_i,\pmb{r}_j)$
with respect to the distribution
\begin{equation}
    P(\{r\}|J,\{\tilde{\sigma}\}) = \frac{e^{\frac{J}{2}\sum_{i\neq j}\tilde{\sigma}_i \tilde{\sigma}_j \Delta(\pmb{r}_i,\pmb{r}_j)}}{\Phi(J,\{\tilde{\sigma}\})} \, .
\end{equation}
We can sample from this distribution using our simulator by fixing the spin configuration to $\{\tilde{\sigma}\}$ and by making only Monte Carlo moves on the polymer degrees of freedom.
So in the end we have
\begin{equation}
    \frac{\partial \ln{\Phi(J,\{\tilde{\sigma}\})}}{\partial J} = \mathbb{E}_{\{r\}|J,\{\tilde{\sigma}\}}\left[ \frac{1}{2}\sum_{i\neq j}\tilde{\sigma}_i \tilde{\sigma}_j \Delta(\pmb{r}_i,\pmb{r}_j)\right] \, ,
\end{equation}
while, for the second derivative, we get
\begin{eqnarray}
    \frac{\partial ^2 \ln{\Phi(J,\{\tilde{\sigma}\})}}{\partial J ^2}
    & = & \nonumber \frac{1}{\Phi(J,\{\tilde{\sigma}\})}\sum_{\{r\} \in\, SAW} \left(\frac{1}{2}\sum_{i\neq j}\tilde{\sigma}_i \tilde{\sigma}_j \Delta(\pmb{r}_i,\pmb{r}_j) \right)^2 e^{\frac{J}{2}\sum_{i\neq j}\tilde{\sigma}_i \tilde{\sigma}_j \Delta(\pmb{r}_i,\pmb{r}_j)} - \nonumber\\
     & & - \left(\frac{1}{\Phi(J,\{\tilde{\sigma}\})}\sum_{r \in\, SAW}\frac{1}{2}\sum_{i\neq j}\tilde{\sigma}_i \tilde{\sigma}_j \Delta(\pmb{r}_i,\pmb{r}_j) e^{\frac{J}{2}\sum_{i\neq j}\tilde{\sigma}_i \tilde{\sigma}_j \Delta(\pmb{r}_i,\pmb{r}_j)}\right)^2 \\
     & = & \mathbb{V}_{r|J,\{\tilde{\sigma}\}}\left[\frac{1}{2}\sum_{i\neq j}\tilde{\sigma}_i \tilde{\sigma}_j \Delta(\pmb{r}_i,\pmb{r}_j)\right] \, . \nonumber
\end{eqnarray}

The major problem with this approximated procedures arises when one has many data points.
In fact, let us suppose
$\{\{\tilde{\sigma}\}_m\}_{m=1}^M$
are M \emph{i.i.d} observations, then the posterior reads:
\begin{equation}
    P(\pmb{\theta}|\{\{\tilde{\sigma}\}_m\}_{m=1}^M) \propto P(\pmb{\theta}) \prod_{m=1}^M P(\{\tilde{\sigma}\}_n|\pmb{\theta})  = \frac{P(\pmb{\theta})}{Z(\pmb{\theta})^M} \prod_{m=1}^M e^{-\mathcal{H}_{\rm ext} ( \{\tilde{\sigma}\}_m | \pmb{w})}\Phi(J,\{\tilde{\sigma}\}_m) \, .
\end{equation}
For the estimation of the log-ratio of the partition functions we just need to multiply $M$ times the quantity obtained for the case of a single data point, whereas for $\Phi(J,\{\tilde{\sigma}\}_m)$ we need to run separate simulations to get different samples (one for each observation).
Although, in principle, this might increase considerably the cost of computing an accurate posterior sampling, for the kind of data we need the number of observations typically ranges from 1 to 3, therefore the increase in the computational burden is expected to be still affordable.

\subsection*{Details about the employed experimental data}
The histone modifications for {\it H3K27me3} and the transcription factor (TF) binding data have been downloaded from the \emph{ENCODE project} database~\cite{encode} for the H1-hESC cell line.
We have considered a variety of different chromosomal regions from human chromosome 2 at different levels of resolution.
Since we work with polymers with fixed number of monomers $N=1000$ (see Section~\ref{sec:ModelHamiltonian} in the main text), the adopted resolutions (see Table~\ref{tab:pearson_corr_histone} in the main text) correspond to chromatin filaments of contour lengths in the range from $5 \cdot 10^5$ to $10^7$ basepairs.

\begin{figure}[h]
    \centering
    \includegraphics[width=\textwidth]{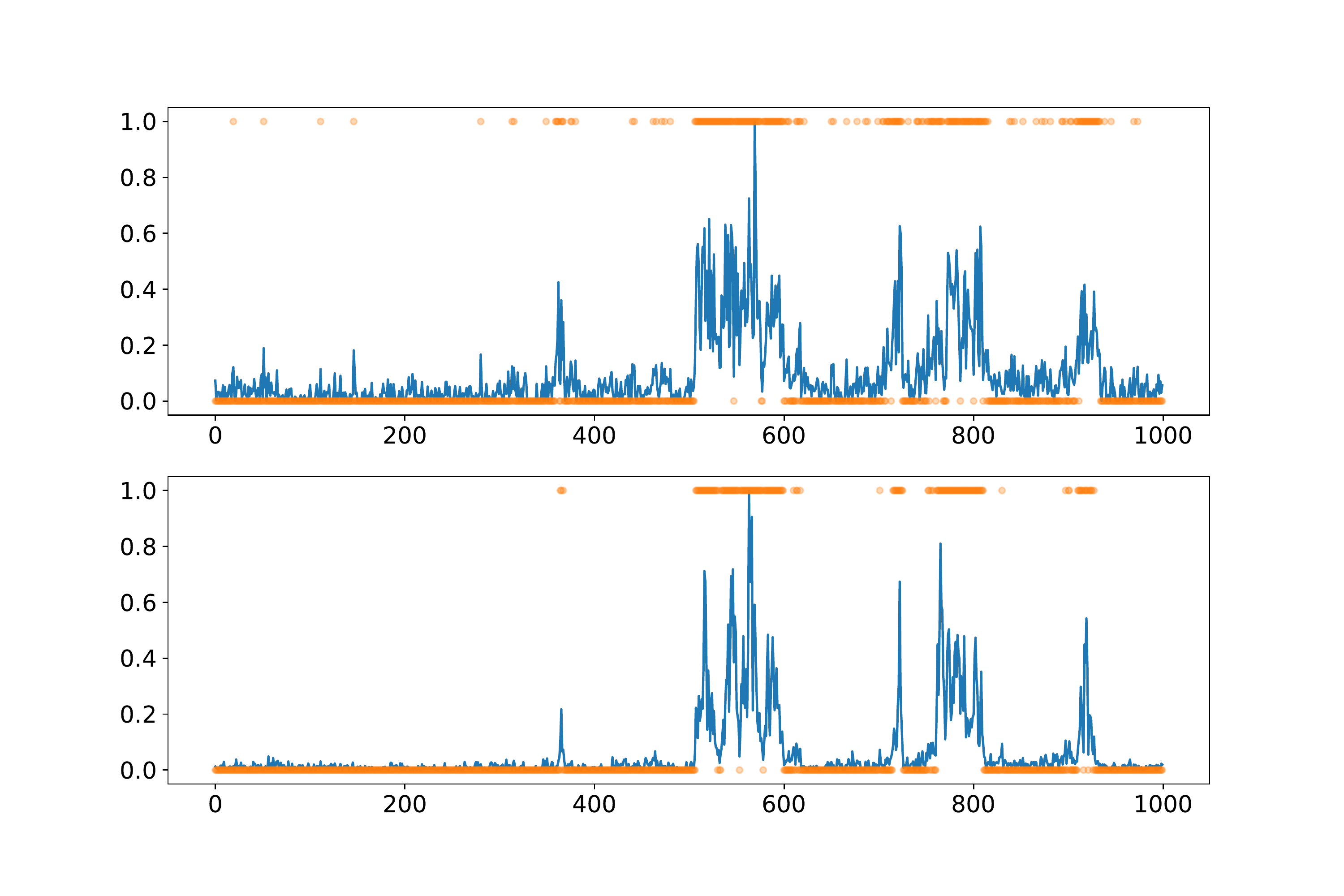}
    \caption{
    The continuous ChIP-seq tracks relative to the histone modification {\it H3K27me3} are discretized in terms of binary spin variables $=0/1$ representing, respectively, the absence/presence of the histone modification.
    The orange dots represent the spin value associated to a specific bin.
    The plots show data corresponding to the region {\it chr2:218,600,000-219,100,000}, discretized in 1000 bins of 500 bps each. 
    }
    \label{fig:Experimental_data}
\end{figure}

The continuous ChIP-seq track for the two biological replicates for the histone modifications have then been converted to discrete valued vectors of 0's and 1's.
Then, each monomer has been assigned value $=0$ or $=1$ depending on whether the signal in the corresponding bin was found below or above the mean signal computed on the entire region (see Fig.~\ref{fig:Experimental_data} for an illustration of the procedure).
The feature matrix $\pmb{F}$ is composed of transcription factor (TF) binding data and CpG density. Accession numbers for all TF binding data can be found in Table~\ref{tab:InputDataDescription} and the CpG density data has been computed from the DNA sequence from the genome assembly GRCh38.
All data went through a denoising pre-processing phase using a lowpass order one Butterworth filter with critical frequency $=0.1$.

The Hi-C matrices used as a comparison for the model generated maps are from the H1-hESC line and can be found on the {\it 4DN} data portal~\cite{4dn} under the accession code 4DNFIQYQWPF5.

\subsection*{Test on synthetic data}
\begin{figure}
    \centering
    \includegraphics[width=\textwidth]{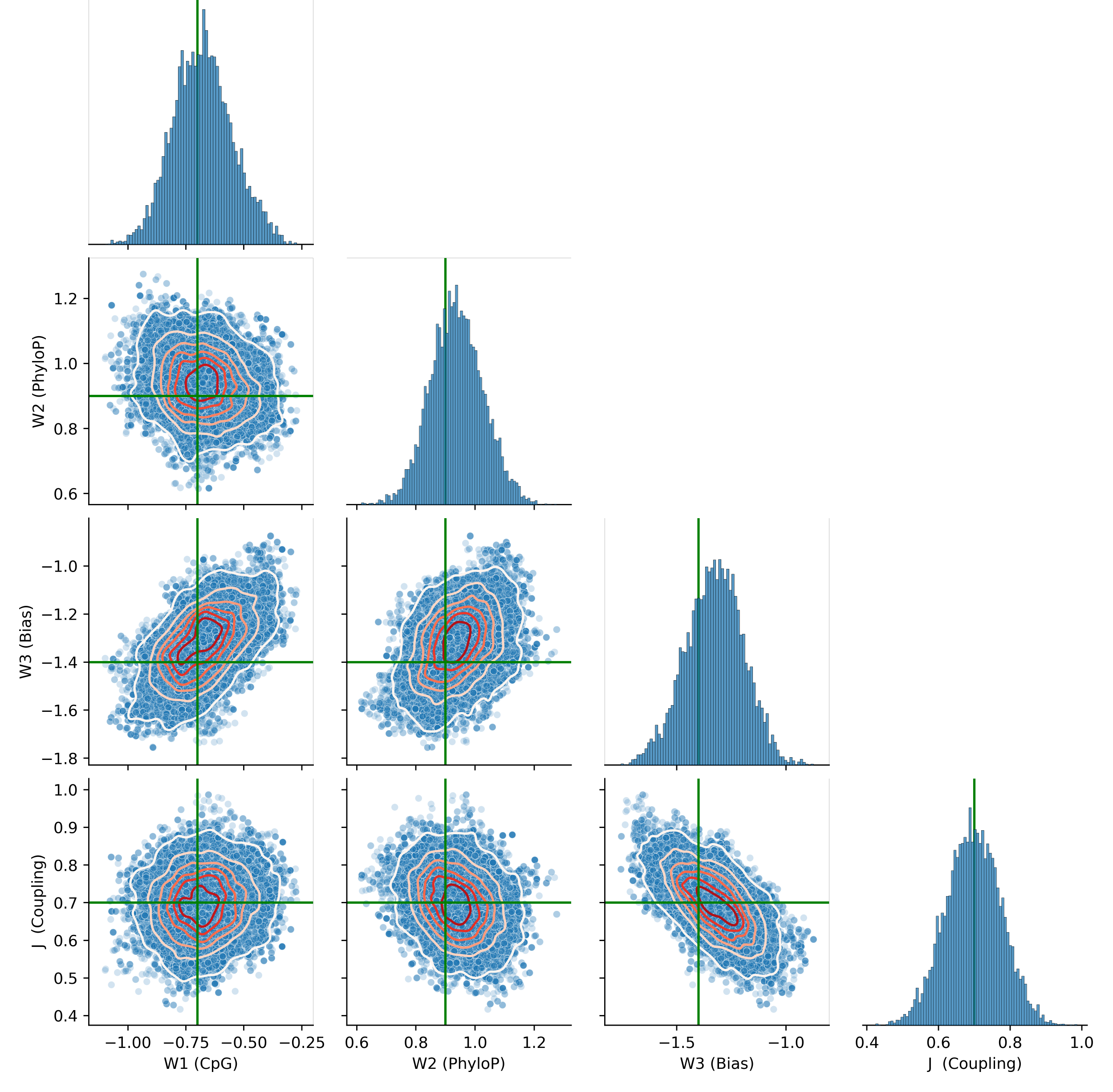}
    \caption{
    Posterior distributions for synthetic input data obtained with our approximate Monte Carlo procedure. 
    Here, 15 spins configurations generated with the simulator were used as input. 
    The green lines identify the {\it ground truth} parameter values used to generate the data.
    }
    \label{fig:synthetic_data_posterior_a}
\end{figure}

To validate our numerical procedure, we test the algorithm first on synthetic data.
Spins configurations are generated using the magnetic polymer simulator with {\it arbitrarily chosen} sets of parameter values as input.
These spin configurations are then used as data points on which to perform posterior inference, see an example of the marginal posterior distributions in Fig.~\ref{fig:synthetic_data_posterior_a}.

\begin{figure}
    \centering
    \includegraphics[width=\textwidth]{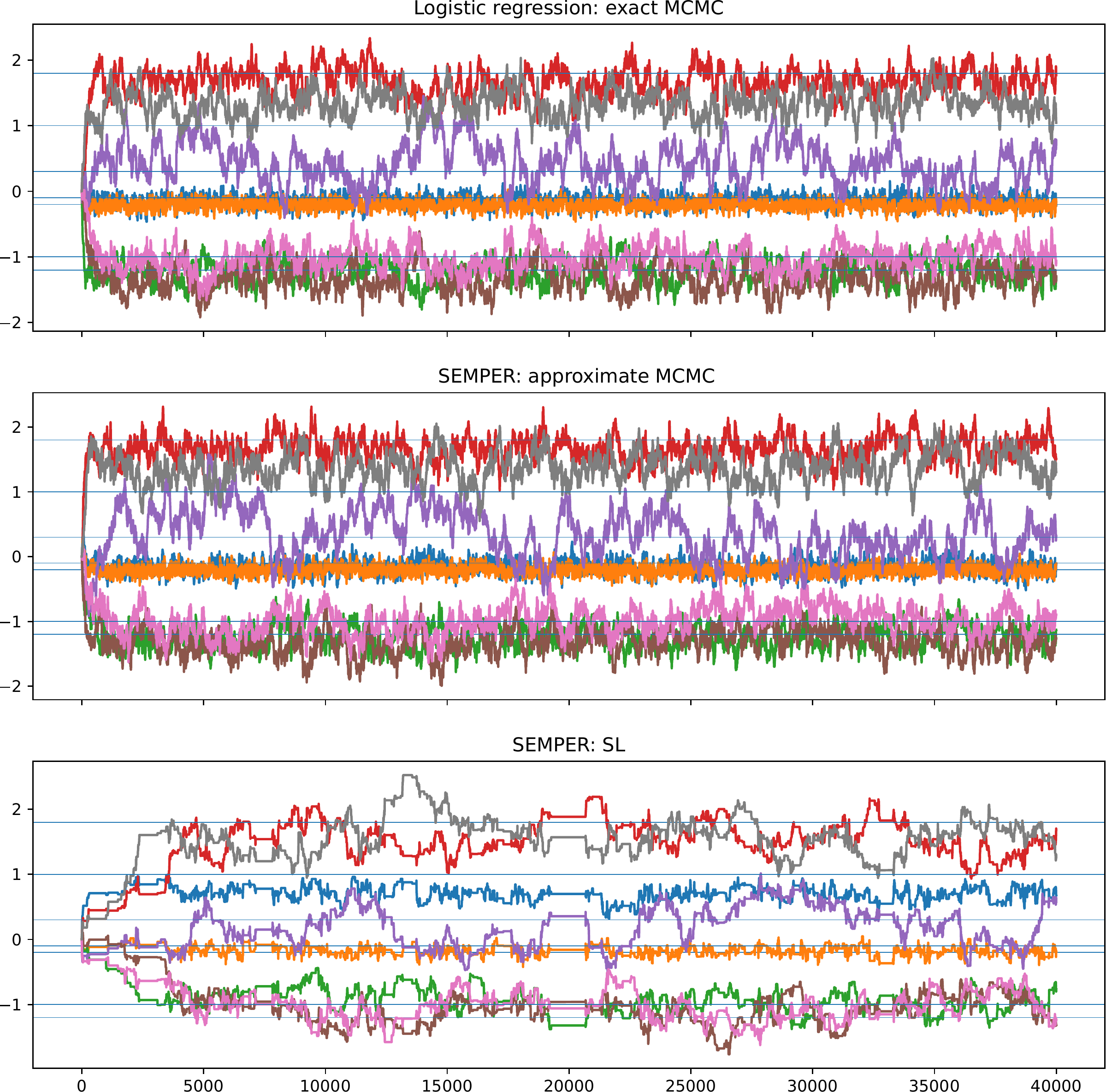}
    \caption{
    MCMC trajectories for posterior sampling from synthetic data generated at fixed $J=0$, using the same set of features employed in the real case study.
    (Top)
    Tracks for the {\it exact} sampling obtained through the logistic regression (LR) procedure.
    (Center)
    Tracks for the approximate Monte Carlo method developed in this work.
    Notice the remarkable agreement with the tracks obtained by exact LR.
    (Bottom)
    Tracks obtained by the likelyhood-free method known as \emph{Synthetic Likelihood} (SL)~\cite{wood10}.
    Notice the poor reproducibility of the tracks, due to the bad mixing of the Markov chains.
    }
    \label{fig:MCMC_trajs_comparison}
\end{figure}

\begin{figure}
    \centering
    \includegraphics[width=\textwidth]{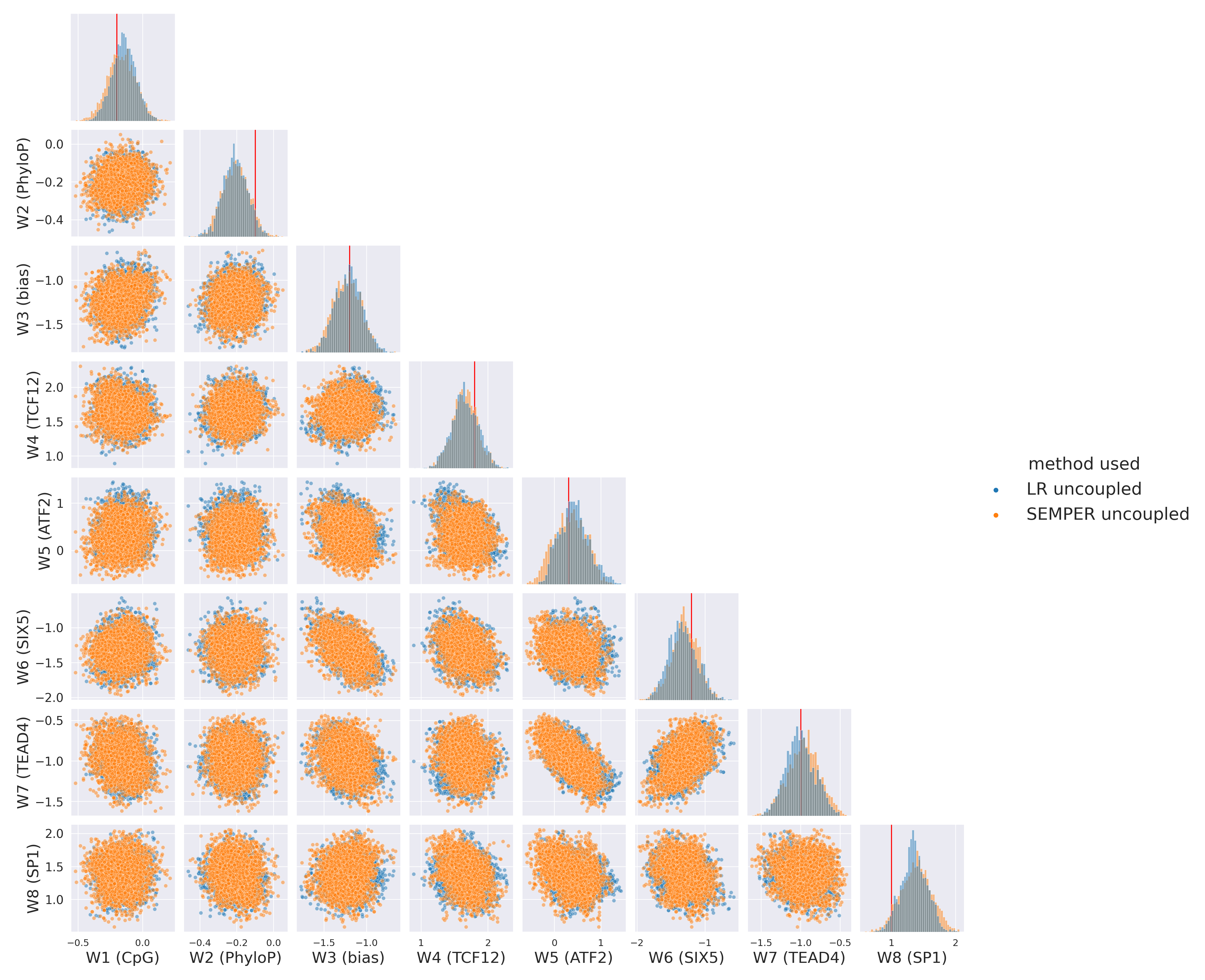}
    \caption{
    Comparison between logistic regression (LR) {\it vs.} SEMPER.
    Histograms and scatter plots for the MCMC samples shown in the top and middle panels of Fig.~\ref{fig:MCMC_trajs_comparison}. 
    }
    \label{fig:synthetic_data_posterior_c}
\end{figure}

\begin{figure}
    \centering
    \includegraphics[width=\textwidth]{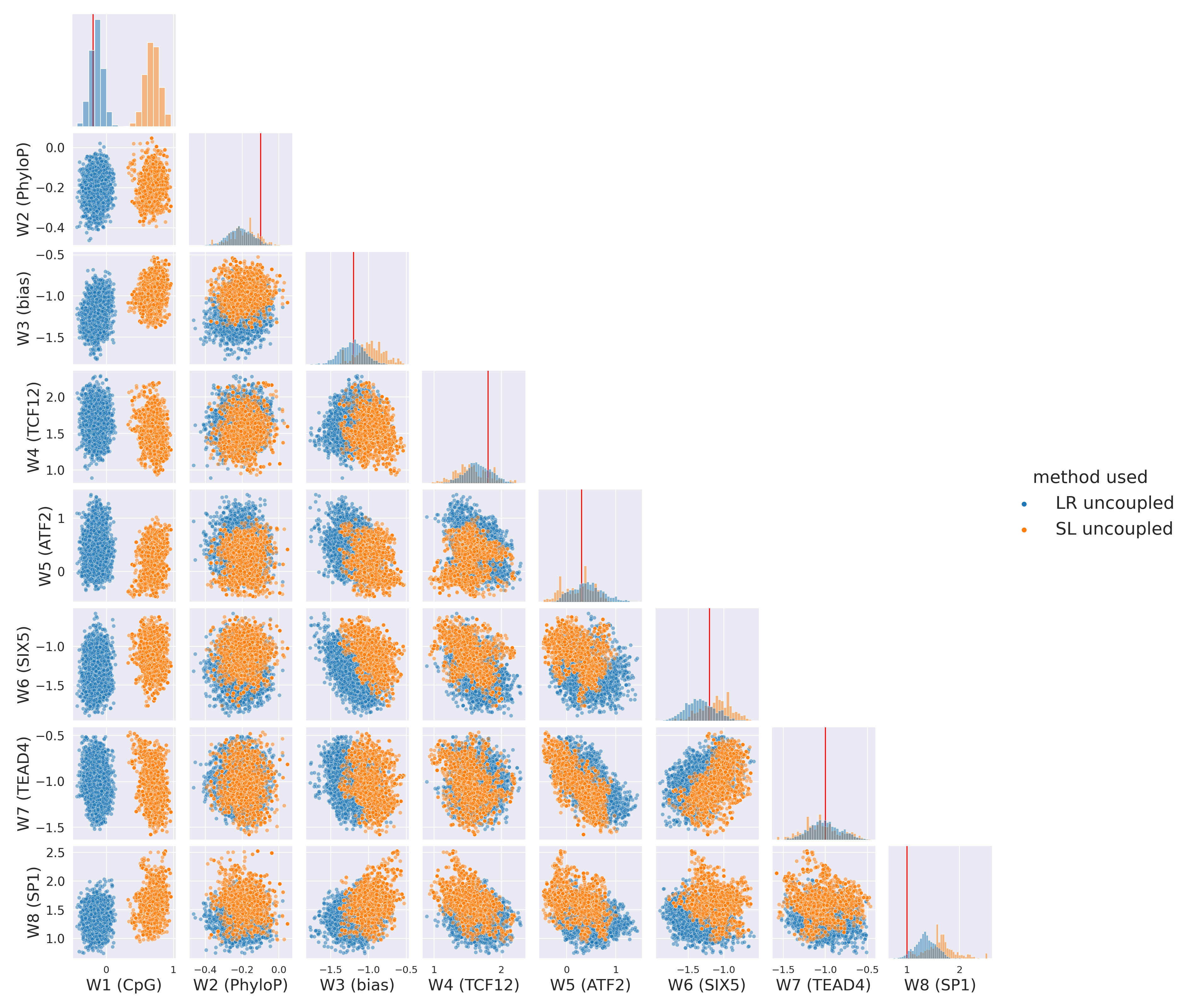}
    \caption{
    Comparison between logistic regression (LR) and synthetic likelihood (SL).
    Histograms and scatter plots for the MCMC samples shown in the middle and bottom panels of Fig.~\ref{fig:MCMC_trajs_comparison}. 
    }
    \label{fig:synthetic_data_posterior_sl}
\end{figure}

Since the {\it true} posterior distributions for the problem at hand are intractable it is non-trivial to assess the goodness of our approximate method.
On the other hand -- as pointed out in the main text -- in the case where the spin-spin coupling $J$ is fixed to zero ({\it i.e.}, the polymer's and spins' degrees of freedom are not coupled) the model reduces to a logistic regression model~\cite{Gelman2013BayesianDA, Benveniste2014} for which, instead, exact posterior sampling is tractable and fast.
One of the reference approximate methods in the field of \emph{likelihood-free inference}~\cite{beaumont02} is the so called \emph{synthetic likelihood} (SL) method~\cite{wood10}. 
To show the quality of our approximate method, we have performed additional numerical experiments in which the synthetic data are generated by setting $J=0$.
Then, posterior samples are computed with the two techniques ({\it i.e.}, the present method {\it vs.} SL), and are compared with the ones obtained with exact sampling of the LR model.
Figs.~\ref{fig:MCMC_trajs_comparison} and~\ref{fig:synthetic_data_posterior_c} demonstrate the excellent accordance between the distributions generated with the exact and our approximate method while, with the same computational cost, the SL method is vastly outperformed, see Figs. \ref{fig:MCMC_trajs_comparison} and~\ref{fig:synthetic_data_posterior_sl}. 

\clearpage

\section{Supplementary Figures and Table}\label{app:figs}

%

\clearpage
\begin{table}[ht]
    \centering
    \begin{tabular}{|c|c|c|}
    \hline
    Measured quantity & Experiment accession number & Data file number \\
    \hline
    H3K27me3 replicate 1 & ENCSR928HYM & ENCFF417VQQ \\
    H3K27me3 replicate 2 & ENCSR928HYM & ENCFF780FNS \\
    CTCF     & ENCSR000AMF & ENCFF332TNJ \\
    H3K4me3  & ENCSR019SQX & ENCFF493QWY \\
    ATF2     & ENCSR000BQU & ENCFF031CBV \\
    SIX5     & ENCSR000BIQ & ENCFF724HLZ \\
    SP1      & ENCSR000BIR & ENCFF684HIL \\
    TCF12    & ENCSR000BIT & ENCFF717UNQ \\
    TEAD4    & ENCSR000BRY & ENCFF307KTY \\
    \hline
    \end{tabular}
    \caption{
    \emph{ENCODE} project accession numbers for the histone modifications and TF binding data used in this work.
    \label{tab:InputDataDescription}
    }
\end{table}
%



\bibliography{apssamp}

\end{document}